\documentclass[journal=jctcce,manuscript=article,layout=traditional]{achemso}
\usepackage{graphicx}
\usepackage{caption}
\usepackage{subcaption}
\usepackage{bm}
\usepackage{braket}
\usepackage[colorlinks=true,breaklinks=true,linkcolor=red,citecolor=magenta,urlcolor=blue]{hyperref}
\usepackage{comment}
\usepackage{amsmath}
\usepackage{amsfonts,amssymb}
\usepackage{multirow}
\usepackage{multicol}

\newcommand{\bts}[1]{{\bf{#1}}}
\newcommand{\scientific}[2]{#1 \cdot 10^{#2}}
\newcommand{\fes}{\mathrm{[2Fe-2S]}}

\usepackage[utf8]{inputenc}
\usepackage[T1]{fontenc}
\usepackage{qcircuit}
\usepackage{float}

\title{Enhancing the accuracy and efficiency of sample-based quantum diagonalization with phaseless auxiliary-field quantum Monte Carlo}
\author{Don Danilov}
\affiliation{Department of Chemistry, Rice University, Houston, TX 77005-1892, USA}
\author{Javier Robledo-Moreno}
\affiliation{IBM Quantum, IBM T.J. Watson Research Center, Yorktown Heights, NY 10598, USA}
\author{Kevin J. Sung}
\affiliation{IBM Quantum, IBM T.J. Watson Research Center, Yorktown Heights, NY 10598, USA}
\author{Mario Motta}
\affiliation{IBM Quantum, IBM T.J. Watson Research Center, Yorktown Heights, NY 10598, USA}
\author{James Shee}
\affiliation{Department of Chemistry, Rice University, Houston, TX 77005-1892, USA}
\alsoaffiliation{Department of Physics and Astronomy, Rice University, Houston, TX 77005-1892, USA}
\email{james.shee@rice.edu}

\begin{document}

\begin{abstract}
Quantum Selected Configuration Interaction (QSCI) and an extended protocol known as Sample-based Quantum Diagonalization (SQD) have emerged as promising algorithms to solve the electronic Schr{\"o}dinger equation with noisy quantum computers.  In QSCI/SQD a quantum circuit is repeatedly prepared on the quantum device, and measured configurations form a subspace of the many-body Hilbert space in which the Hamiltonian is diagonalized classically.
For the dissociation of N$_2$ and a model $\fes$ cluster (correlating 10 electrons in 26 orbitals and 30 electrons in 20 orbitals, respectively) we show that a non-perturbative stochastic approach, phaseless auxiliary-field quantum Monte Carlo (ph-AFQMC), using truncated SQD trial wavefunctions obtained from quantum hardware can recover a substantial amount (e.g., $\mathcal{O}$(100) mHa) of correlation energy.  %Extrapolation of the ph-AFQMC energy versus the energy variance of the SQD trial wavefunctions has the potential to further improve the energy accuracy.  
This hybrid quantum-classical combination has the potential to greatly reduce the sampling burden placed on the QSCI/SQD procedure, and is a compelling alternative to recently proposed hybrid ph-AFQMC algorithms that require VQE-based tomography.
\end{abstract}

\maketitle

\section{Introduction}

The concerted use of classical and quantum computers has emerged as a promising strategy to tackle problems in many-body simulation and other fields of science~\cite{cao2019quantum,bauer2020quantum,mcardle2020quantum,cerezo2021variational,motta2022emerging}.
In particular, hybrid quantum-classical algorithms~\cite{mcclean2016theory} and the computational platforms to implement them, called quantum-centric supercomputers~\cite{alexeev2024quantum}, are increasingly being used to perform electronic structure simulations.  For example, a quantum computer can prepare active-space multireference wavefunctions while a classical computer performs pre-, peri- and post-processing operations to mitigate errors occurring on quantum devices~\cite{} and/or to capture dynamical electron correlation effects~\cite{huang2023leveraging,fitzpatrick2024quantum,motta2020quantum,erhart2024coupled,tammaro2023n,khinevich2025enhancing}.
Another scheme involves the use of wavefunctions prepared by a quantum computer as so-called trial wavefunctions~\cite{huggins2022unbiasing} in phaseless auxiliary-field quantum Monte Carlo (ph-AFQMC) calculations~\cite{zhang2003quantum}.

AFQMC approximates ground-state wavefunctions by propagating an initial state (typically, though not exclusively, a Slater determinant) in imaginary time,
and does so by mapping the imaginary-time evolution onto a stochastic process that samples the manifold of non-orthogonal Slater determinants~\cite{stratonovich_method_1957,hubbard_calculation_1959}.
With unconstrained random walks, AFQMC suffers from a sign or phase problem, i.e. an exponential growth of statistical uncertainties on observable properties with system size and imaginary time.
For Hamiltonians with the Coulomb interaction, the phase problem is controlled by the phaseless constraint~\cite{zhang2003quantum}, in which a trial wavefunction is used to introduce a drift term and a branching factor in the random walk, enabling the removal of problematic samples that have acquired an excess of complex phase (that causes uncontrolled statistical fluctuations).
However, controlling the phase problem biases computed observables, to an extent determined by the quality of the trial wavefunction, and requires numerous evaluations of overlaps and Hamiltonian matrix elements (local energies) between the trial wavefunction and walker determinants.

The first study using wavefunctions obtained from quantum computation as ph-AFQMC trials was made by Huggins et al~\cite{huggins2022unbiasing}, employing the variational quantum eigensolver (VQE) method~\cite{peruzzo2014variational} to variationally optimize an ansatz, and using the optimized wavefunction as the ph-AFQMC trial.
The extraction of information from the quantum computer to evaluate overlaps and local energies was based on shadow tomography~\cite{aaronson2018shadow,zhao2021fermionic} to learn a classical representation of the quantum trial.
This method, later refined by Huang et al and Zhao et al employing matchgate shadows~\cite{huang2024evaluating, zhao2025quantumclassicalauxiliaryfieldquantum} and Kiser et al through the contextual subspace AFQMC method~\cite{kiser2024contextual}, has been argued to be a promising candidate to tackle important chemical problems using classical and quantum computers in concert~\cite{amsler2023classical,kiser2024classical}.

The above approaches to hybrid quantum-classical ph-AFQMC offer several benefits, including the ability to significantly unbias ph-AFQMC estimates via the use of non-linear, unitary trial wavefunctions.  In addition, the scheme allows for the recovery of dynamical electron correlation due to excitations involving electrons and orbitals outside of the active space~\cite{khinevich2025enhancing}.
However, employing shadow tomography to obtain overlaps and local energies poses several challenges. Firstly, statistical uncertainties affecting these quantities need to be resolved with high precision, which results in a high computational cost~\cite{mazzola2022exponential}.
Secondly, while quantum-computing trials have the potential to improve the phaseless constraint, the tomography-based procedure requires VQE, which on near-term quantum computers is severely limited by device noise in terms of ansatz flexibility, convergence, and optimizability.

Recently, an alternative to VQE has emerged: the quantum selected configuration interaction~\cite{kanno2023quantum} (QSCI) and its extension known as sample-based diagonalization (SQD)~\cite{robledo2024chemistry}, which led to relatively large-scale demonstrations involving circuits of 42 to 77 qubits~\cite{robledo2024chemistry,barison2024quantum,kaliakin2024accurate,shajan2024towards,kaliakin2025implicit}.
SQD is a form of selected configuration interaction in which configurations (i.e. Slater determinants) are obtained by sampling a quantum circuit; error mitigation at the level of individual samples is achieved by enforcing conservation of particle number and other molecular symmetries~\cite{robledo2024chemistry}.
An SQD calculation returns a linear combination of orthogonal Slater determinants with coefficients determined by diagonalization on a classical computer. This functional form makes such variational SQD wavefunctions very convenient for use as AFQMC trial wavefunctions, bypassing the need for shadow tomography (simply because the trial wavefunction is in a form that requires no classical learning).
The extent to which SQD wavefunctions can benefit AFQMC by mitigating the bias from the phaseless constraint is still an open problem, due to the impact of quantum noise in SQD calculations and because the nature of optimal wavefunctions for configuration sampling is not yet understood nor established~\cite{reinholdt2025exposing}. Nevertheless, the use of AFQMC post-processing is a compelling way to recover missing correlation energy from the variational SQD wavefunction and to offer a more balanced treatment of static and dynamical correlations.

\section{Methods}

\begin{figure}[H]
\centering
\includegraphics[width=\textwidth]{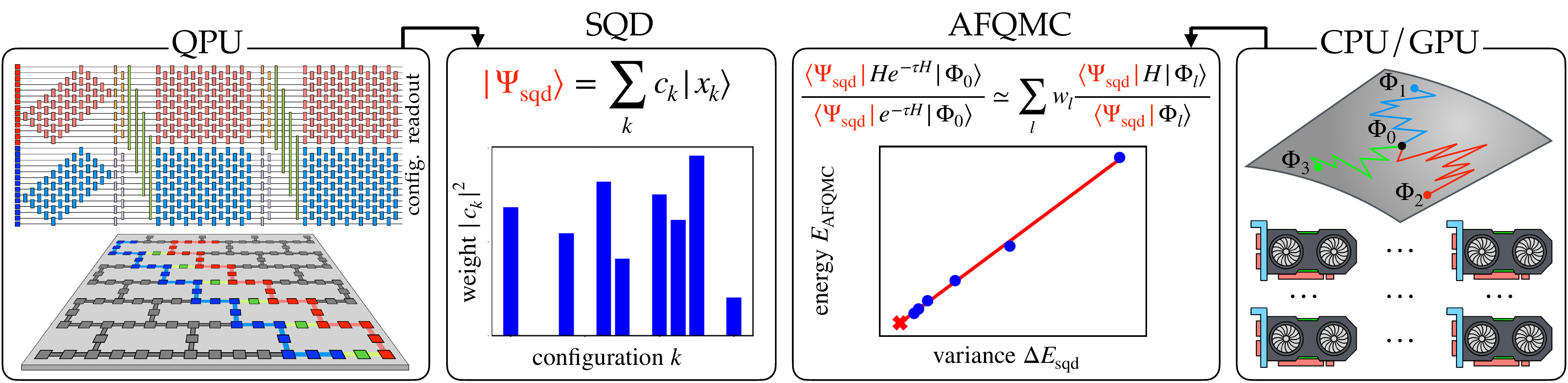}
\caption{Schematic representation of the workflow introduced in this study. An SQD trial wavefunction (second panel) is produced by measuring configurations over an LUCJ quantum circuit (first panel, top) executed on a superconducting quantum processor (first panel, bottom), and performing a configuration recovery and diagonalization on a classical computer. The SQD wavefunction, a linear combination of Slater determinants (second panel, histogram), is used as the trial in a ph-AFQMC calculation (third panel). The ph-AFQMC method maps the imaginary-time evolution onto a random walk in the space of Slater determinants (fourth panel, top), accelerated with GPUs (fourth panel, bottom), with the trial wavefunction controlling the sign/phase problem. 
We compute ph-AFQMC energies for various SQD trials, and perform energy-variance extrapolations (third panel).
}
\label{fig:schema}
\end{figure}

The methods explored in this study are schematically represented in Fig.~\ref{fig:schema}.

\subsection{QSCI, SQD, and the LUCJ ansatz}

QSCI \cite{kanno2023quantum} is a hybrid quantum-classical algorithm in which a quantum computer is used to generate electronic configurations by sampling them from a quantum circuit, and then those configurations are used to form a subspace in which to project and diagonalize the electronic Hamiltonian. SQD \cite{robledo2024chemistry} is an extension of QSCI designed for noisy quantum computers. In SQD, an error mitigation procedure is applied to sampled configurations that violate known symmetries of the system in order to recover valid configurations that can be used to form the subspace for diagonalization.

The performance of QSCI (and by extension, SQD) depends greatly on the quantum circuit used to sample the electronic configurations. Specifically, the circuit should be able to generate configurations on which the target wavefunction has significant support (i.e., CI weight). Similarly to VQE \cite{peruzzo2014variational}, QSCI uses a circuit ansatz whose parameters can be optimized to minimize energy. While VQE requires measuring the expectation value of the Hamiltonian to high precision, which can incur a prohibitive sampling overhead on the quantum computer~\cite{wecker2015progress}, QSCI directly uses quantum samples for its classical diagonalization.

In this work, we consider SQD with samples drawn from the local unitary cluster Jastrow (LUCJ) circuit ansatz \cite{motta2023bridging,motta2024quantum}. The LUCJ ansatz is a variant of the unitary cluster Jastrow (UCJ) ansatz \cite{matsuzawa2020jastrow} tailored for quantum processors with limited qubit connectivity, such as a square or more restricted lattice. We consider the single-layer version of the ansatz with a final orbital rotation, in which case the ansatz has the form
\begin{equation}
\label{eq:lucj}
| \Phi_{\mathrm{qc}} \rangle = e^{-\hat{K}_2} e^{\hat{K}_1} e^{i\hat{J}_{1}} e^{-\hat{K}_{1}} | {\bf{x}}_{\mathrm{RHF}} \rangle \;,
\end{equation}
where $| {\bf{x}}_{\mathrm{RHF}} \rangle$ is the restricted Hartree-Fock (RHF) state, ${\hat{K}}_1$ and ${\hat{K}}_2$ are one-body operators,
and ${\hat{J}}_1 = \sum_{p\sigma r\tau} J_{p\sigma,r\tau} \hat{n}_{p\sigma} \hat{n}_{r\tau}$ is a density-density operator.

We consider a few different ways of obtaining parameters for the ansatz:
\begin{itemize}
    \item Truncated factorization of $t_2$ amplitudes obtained from a coupled-cluster, singles and doubles (CCSD) calculation \cite{motta2023bridging}.  We refer to the subsequent SQD as SQD(CCSD).
    \item Numerical optimization of the parameters to minimize the SQD energy using the classical optimization algorithm COBYQA \cite{rago_thesis,razh_cobyqa}. We performed this optimization using classical simulation, but in the future it could be performed on a quantum computer. We refer to this as SQD(OPT).
    \item Numerical optimization of the parameters to minimize the SQD energy incorporating knowledge of a wavefunction obtained from an SCI calculation, using a procedure described in the Supplementary Material of Ref.~\citenum{robledo2024chemistry}. Specifically, we (i) perform an SCI calculation with a conservative truncation threshold of $\varepsilon_1 = \scientific{5}{-5}$, (ii) optimize the Kullback–Leibler (KL) divergence between the probability distribution of samples drawn from the LUCJ circuit from the SCI wavefunction as a function of the parameters in the LUCJ wavefunction, and (iii) further optimize the parameters of the LUCJ wavefunction to minimize the SQD energy using differential evolution \cite{storn1997differential}. This optimization can only be performed using classical simulation, and we use it to probe the expressiveness of the LUCJ ansatz. The optimization of the KL divergence in step (ii) is performed as a two-step process. In the first step we set $J_{p\sigma, r\tau} = 0$, resulting in a circuit that represents a Slater determinant. Such circuit can be optimized with a cost that is polynomial in the number of electrons. In the second step we allow the $J_{p\sigma, r\tau}$ to be jointly optimized with the parameters of the one-body operators of the LUCJ ansatz. We refer to this as SQD(ovlpOPT). %\js{Clarify, if possible.}
\end{itemize}

\subsection{ph-AFQMC}

ph-AFQMC\cite{zhang2003quantum,motta_ab_2018,lee_twenty_2022,shee_potentially_2023}  performs open-ended random walks in a space of non-orthogonal Slater determinants constrained by a ``trial wavefunction'', which affords low-polynomial scaling computational cost with system size at the expense of a systematically-improvable bias.  
For a given trial wavefunction, ph-AFQMC exhibits low-polynomial scaling with system size.  Recent algorithms have enabled substantial reductions in the scaling prefactor, e.g., correlated sampling algorithms for energy differences\cite{shee2017chemical,chen2023algorithm} orbital localization schemes,\cite{weber2022localized,kurian2023toward} and optimized implementations (of, in principle, near-perfect parallel efficiency) on graphical processing units (GPUs).\cite{shee2018phaseless,malone2020accelerating,jiang2024improved} 
Aside from energies, ph-AFQMC has been shown capable of obtaining gradients and response properties by back-propagation\cite{motta2017computation,motta2018communication} and automatic differentiation\cite{mahajan2023response} approaches.
Extension of ph-AFQMC to excited states have also been documented.\cite{ma_excited_2013}

ph-AFQMC thermochemical predictions with multi-determinant trial wavefunctions have been shown capable of achieving $\sim$1 kcal/mol accuracy vs available experiments for transition metal compounds ranging from atoms\cite{shee2018phaseless} and diatomics\cite{shee2019achieving,mahajan2024beyond} to coordination complexes\cite{rudshteyn2020predicting} and metallocenes.\cite{rudshteyn2022calculation}   
The method has shown promise in predicting the relative energies between states of different spin multiplicity\cite{shee2019singlet,weber2021silico,lee2020utilizing} and can resolve extremely small energy scales.\cite{hao2018accurate,upadhyay2020role}  High accuracy for main group molecules has also been demonstrated.\cite{sukurma2023benchmark,lee_twenty_2022,mahajan2024beyond}

As is important in the regime of strong correlation,\cite{ganoe_notion_2024} the ph-AFQMC method is non-perturbative.  The bias arising from the phaseless approximation depends on the quality of the trial wavefunction, as do properties such as size-consistency and size-extensivity.  In the limit of an exact trial, ph-AFQMC is exact. In practice, one strategy is to converge ph-AFQMC energies with respect to the quality of the trial wavefunction; this has been demonstrated with trials of SCI\cite{holmes2016heat,neugebauer2023toward,mahajan2022selected} or matrix product state\cite{jiang2025unbiasing} forms.  Alternatively, symmetry-broken single-determinant trials, e.g., from generalized Hartree-Fock\cite{danilov_capturing_2024}, appear promising and can be scaled to large molecules and materials. 

\subsection{Computational Details}

\paragraph{Sample-based quantum diagonalization:}

For wavefunctions obtained from hardware experiments, we reused the data obtained in Ref.~\citenum{robledo2024chemistry}, whose details we review here. The experiments were executed on the ibm$\_$torino Heron quantum processor. For N$_2$, 10$^5$ samples were taken from the quantum processor. For $\fes$, $\scientific{2.4567}{6}$ samples were taken. We executed the quantum circuits using Version 1 of the Qiskit \cite{Qiskit} Runtime Sampler Primitive, with readout error mitigation and dynamical decoupling enabled. To postprocess the quantum samples and perform the diagonalization, we used the SQD Qiskit addon \cite{sqd_addon}.

\paragraph{LUCJ ansatz:}

We used ffsim \cite{ffsim} to construct the LUCJ ansatz and simulate it numerically. To obtain LUCJ ansatz parameters numerically optimized for the VQE or QSCI energy, we used the COBYQA \cite{rago_thesis,razh_cobyqa} optimizer, with the initial guess taken to be the truncated CCSD parameters. We used the implementation of COBYQA in SciPy \cite{scipy}. We used the solution obtained by the optimization with a limit of 1000 iterations, even though in all cases the optimization did not converge before reaching that limit. To obtain random parameters, we sampled entries of $\hat{K}_1$, $\hat{K}_2$, and $\hat{J}_{1}$ uniformly at random from the interval (-10, 10).

\paragraph{ph-AFQMC:}  After producing SQD wavefunctions of the form $| \Psi_{\mathrm{sqd}} \rangle = \sum_{i=0}^{d-1} c_i | \bts{x}_i \rangle$ (where, without loss of generality, we can assume that the coefficients $c_i$ are in decreasing order of weight $w_i = |c_i|^2$), we employ them as trial wavefunctions in AFQMC calculations. For the purpose of understanding the performance of AFQMC vis-\`a-vis the quality of the trial wavefunction, for each system under study we consider a single SQD wavefunction and truncate it, retaining the $n_w$ highest-weight coefficients such that $\sum_{i=0}^{n_w} w_i = w$, where $w$ ranges between 50\% and 99.5\%.  %In the ph-AFQMC calculations with experimentally-derived SQD trial wavefunctions for [2Fe-2S] the largest trial had 35535 determinants; in the optimized [2Fe-2S] calculation the largest trial had  116998 determinants.

We used an imaginary time discretization of $\Delta\tau = 0.005$ Ha$^{-1}$ and a Cholesky cutoff of $10^{-12}$. We have used 20480 walkers for the experiment-derived N$_2$/cc-pvdz and $\fes$ calculations and 2048 walkers in all other systems. All trajectories were run for 1000 blocks (each block consisting of 20 time-steps).
Walkers orbitals were re-orthonormalized every 2 steps and energy measurement and population control was performed every 20 steps. 
We performed re-blocking analysis\cite{reblockana} using the pyblock suite. The data reported throughout is that of the optimal statistical block with the lowest energy.

\paragraph{Active spaces:}
We study (i) the dissociation of the nitrogen molecule (N$_2$) and (ii) the ground-state of a methyl-capped $\fes$ cluster. 
For the dissociation of N$_2$, we employ the 6-31G and cc-pVDZ basis sets with the standard frozen-core approximation.  We use the PySCF~\cite{sun2018pyscf,sun2020recent} software to perform RHF calculations and enforce double occupation of the two lowest-energy molecular orbitals corresponding to bonding and anti-bonding linear combinations of N $1s$ orbitals, using standard functions from the mcscf and tools.fcidump libraries of PySCF. For the $\fes$ cluster, we employ the (30e,20o) active space proposed by Sharma et al~\cite{sharma2014low}.

\paragraph{Energy variance calculations:}
In general, SQD calculations do not produce exact eigenfunctions of the Hamiltonian. In practice, it is difficult to analyze the relation between the number and nature of the configurations in the SQD wavefunction and the systematic deviation between the SQD and ground-state energies. To characterize these biases and produce a more accurate estimate of the ground-state energy, we use an energy-variance extrapolation. 
It is known~\cite{kashima2001path} that the difference $\delta E = \langle \Psi_{\mathrm{sqd}} | \hat{H} | \Psi_{\mathrm{sqd}} \rangle - \langle \Psi_{\mathrm{gs}} | \hat{H} | \Psi_{\mathrm{gs}} \rangle$ between the expectation value of $\hat{H}$ over the SQD state and the ground-state energy vanishes linearly as a function of the energy variance $\Delta E_{\mathrm{sqd}} = \frac{ \langle \Psi_{\mathrm{sqd}} | \hat{H}^2 | \Psi_{\mathrm{sqd}} \rangle -\langle \Psi_{\mathrm{sqd}} | \hat{H} | \Psi_{\mathrm{sqd}} \rangle^2}{ \langle \Psi_{\mathrm{sqd}} | \hat{H} | \Psi_{\mathrm{sqd}} \rangle^2}$, i.e. $\delta E \propto \Delta E_{\mathrm{sqd}}$, providing a simple and effective extrapolation procedure. 

In this work, we perform an additional extrapolation, on ph-AFQMC energies: we assume a proportionality relation $E_{\mathrm{ph-AFQMC-SQD}} - \langle \Psi_{\mathrm{gs}} | \hat{H} | \Psi_{\mathrm{gs}} \rangle \propto \Delta E_{\mathrm{sqd}}$, and conduct a linear extrapolation of AFQMC energies computed with different SQD trial wavefunctions, with respect to the trial wavefunction variance, to the zero-trial-variance limit.

\paragraph{Heatbath Configuration Interaction (HCI):} 
HCI constructs an approximation of the molecular wavefunction by iteratively growing an orthogonal CI expansion where configurations are selected based on some criterion; for the purpose of this paper we restrict ourselves to the algorithm where a new candidate configuration $\ket{\bts{x}_i}$ is included in the next iteration of the expansion if $c_j\braket{\bts{x}_j |\hat{H}|\bts{x}_i}$ is larger than some user-defined threshold (referred to as $\epsilon_1$) for any of the $\bra{\bts{x}_j}$ in the current expansion. To generate HCI trials we have used the Dice code~\cite{dice1,dice2} version 1.0.

\section{Results}

\subsection{N$_2$}

\begin{figure}[H]
\centering
\includegraphics[width=0.48\textwidth]{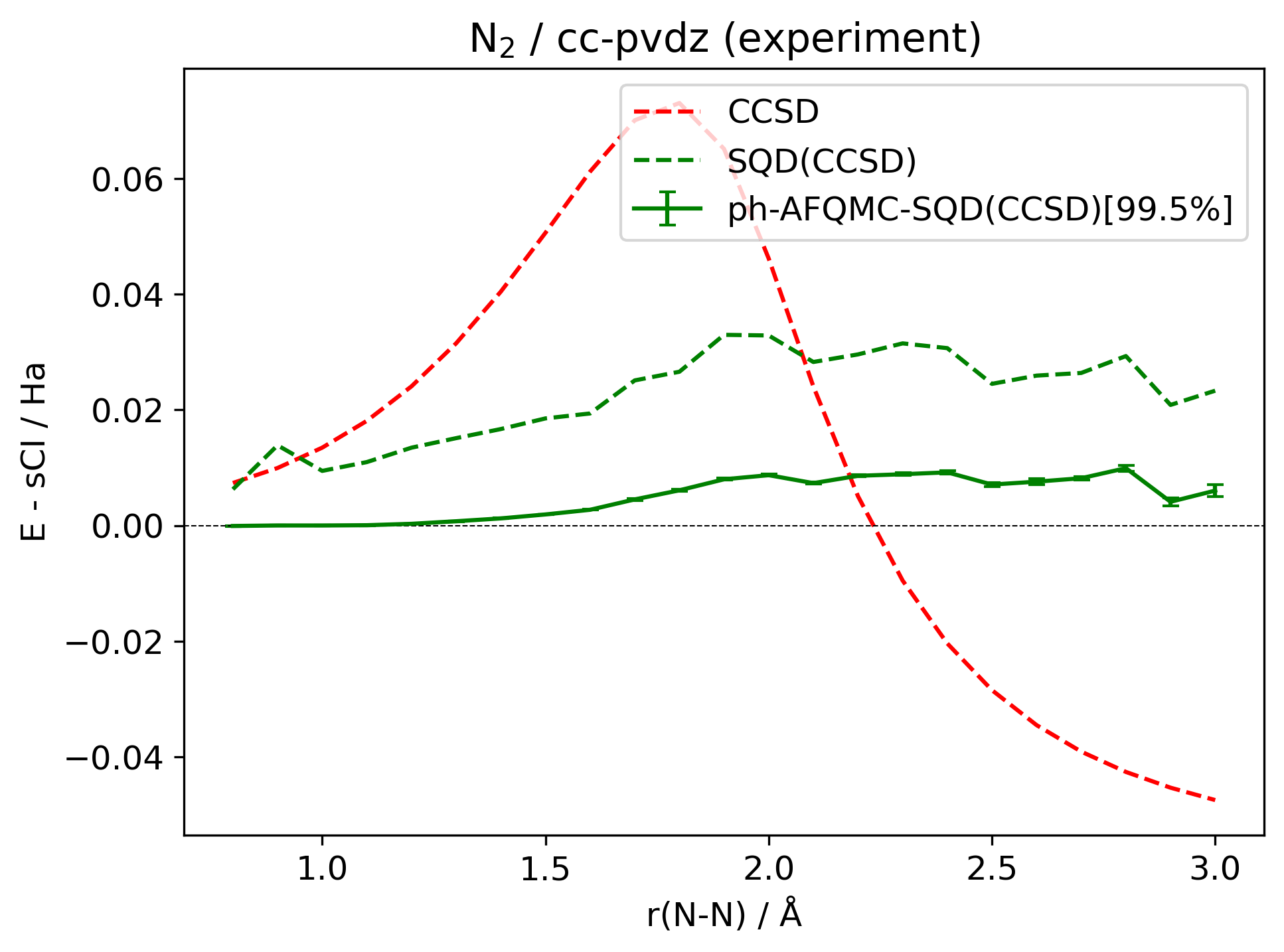}
\includegraphics[width=0.48\textwidth]{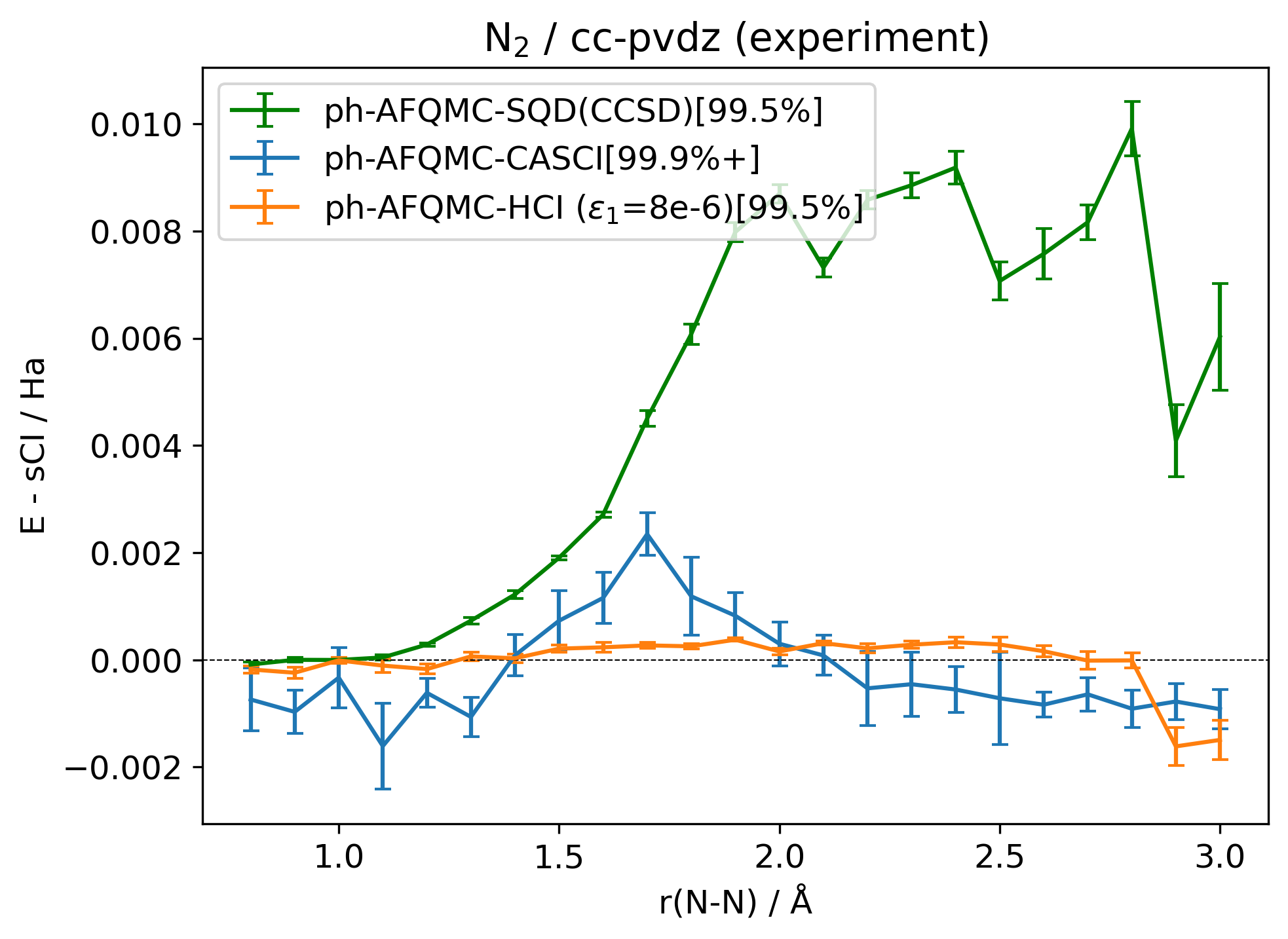}
\caption{Energy errors with respect to converged selected CI, for various methods; CCSD, SQD initialized from a single-layer LUCJ (with SVD'd CCSD parameters) at subspace dimension of 16 million) carried out on the quantum device, and ph-AFQMC with 99.5\% truncated SQD trial wavefunctions (left) and ph-AFQMC using a the larger subspace SQD, CASCI (in a 10e8o active space) and HCI multi-slater determinant trials (right)} 
\label{fig:n2exp}
\end{figure}

Stretching the N$_2$ molecule is a paradigmatic way to access the strongly correlated regime.  The left panel of Figure \ref{fig:n2exp} shows data along the dissociation curve in the cc-pVDZ basis set with a frozen-core approximation resulting in 10 electrons and 26 spatial orbitals, relative to converged selected CI energies in the same space.  CCSD, while accurate at equilibrium, undercorrelates and then overcorrelates as the bond is stretched. The SQD(CCSD) energies are obtained from experiments on real hardware in which samples were drawn from a single-layer LUCJ circuit with the unoptimized parameters implied by a classical CCSD calculation.  A feature of SQD is its variationality, despite hardware noise; however, the energy errors relative to the benchmark range from 6.3 mHa to 33.0 mHa with an average error of 22.2mHa.

%At bond lengths less than 1.7\AA, ph-AFQMC-SQD is very close to the exact total energy.  Notably, for larger bond lengths the error increases to as much as 41.8 mHa at 2.2\AA.  Taking a closer look at the ph-AFQMC-SQD energy vs imaginary-time trajectories (the right panel of Figure \ref{fig:n2exp} corresponds to the 2.2\AA \ bond length) reveals a non-monotonic behavior that indicates that the trial wavefunction used is of suboptimal quality.  We hypothesize that the hardware-derived SQD wavefunctions beyond 1.7\AA \ have some excited singlet state character (which could be quantified as an artificially large overlap with one or more exact eigenstate other than the ground-state).  This would explain the unexpected drift up in the ph-AFQMC-SQD energy trajectories, and requires further investigation.

When the SQD wavefunction is truncated such that 99.5\% of the CI weight is retained, and used as a trial wavefunction for classical ph-AFQMC, the energies are substantially improved across the dissociation curve. In the right panel of Fig. \ref{fig:n2exp}, the ph-AFQMC-SQD(CCSD) energies are compared with results from the use of state-of-the-art multiconfigurational trial wavefunctions generated on classical computers.  Specifically, we consider CASCI wavefunctions with an active space of 10 electrons and 8 orbitals, and HCI expansions using $\epsilon_1 = 8 \cdot  10^{-6}$.  These purely classical trials lead to near-exact ph-AFQMC energies, though the SQD(CCSD) trial ph-AFQMC energies are at most 10 mHa above the reference values across the dissociation curve.  We emphasize that the quality of an SQD wavefunction depends critically on the circuit that provides the probability distribution from which samples are drawn.  In what follows we will show that the CCSD initialization procedure employed in SQD(CCSD) is suboptimal, and that both the SQD and ph-AFQMC-SQD energies can be substantially improved.

\begin{figure}[H]
\centering
\includegraphics[width=0.75\textwidth]{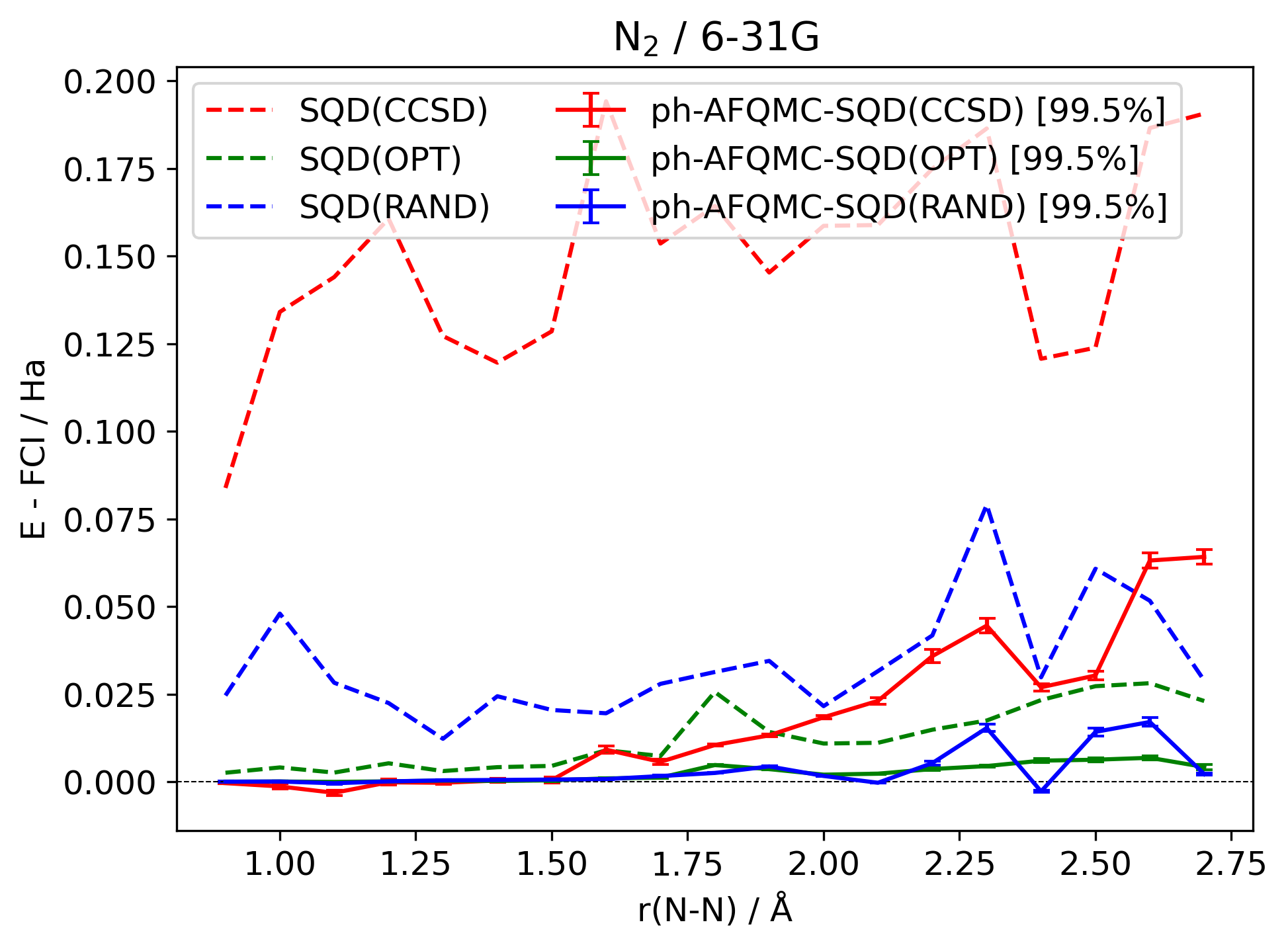}
\caption{Energy errors in the 6-31G basis set with respect to FCI for SQD approaches initialized from a single-layer LUCJ circuit with parameters from:  CCSD amplitudes, variational minimization of the SQD energy, and random sampling. }
\label{fig:n2opt}
\end{figure}

In Figure \ref{fig:n2opt} we use a smaller basis set (6-31G, in which FCI is computationally feasible), and explore the dependence of the SQD and ph-AFQMC-SQD energies on the initial LUCJ wavefunction parameters.  Using a single-layer wavefunction ansatz, as before, we obtain LUCJ parameters from decomposed CCSD amplitudes, variational optimization with respect to the SQD energy, or a procedure based on random selection. 
We find that the SQD energies improve (relative to FCI values) going from SQD(CCSD) to SQD(RAND) to SQD(OPT).
In all cases, subsequent ph-AFQMC lowers the energy substantially vs SQD alone.  This is most dramatically seen in the SQD(CCSD) case, where ph-AFQMC recovers between roughly 80-200 mHa of correlation energy missed by SQD.  As alluded to above, using more sophisticated trial wavefunctions than SQD(CCSD) -- e.g., those obtained from the SQD(OPT) protocol -- notably improves the ph-AFQMC energies, with an average error of 2.5 mHa and a maximum error of 6.9 mHa relative to FCI. 

Rather surprisingly, the accuracy of ph-AFQMC with SQD trials obtained from random LUCJ parameter generation is remarkable; there is virtually no error in the energy until after 2.1 \AA. %, after which the error is at most some 20 mHa; 
%the ground state of N$_2$ acquires multireference character as $R$ increases, and the combination of device noise and circuit parametrization challenges the accuracy of SQD in this regime.
In this case, the CCSD wavefunction when represented in the space of Slater determinants has most of its weight on the HF configuration.  Such peaked distributions lead to suboptimal configuration sampling (which is problematic given the finite number of samples drawn in practical SQD procedures).  In contrast, sampling the LUCJ parameters randomly likely mimics a more uniform distribution, which leads to improved SQD sampling efficiency and, therefore, a more accurate ph-AFQMC-SQD energy.  Note that  this level of accuracy of the SQD(RAND) protocol is not generally transferrable beyond small system sizes (with the possible, and interesting, exception of maximally frustrated many-body systems).

\subsection{$\fes$ model cluster}

\begin{figure}[H]
\centering
\includegraphics[width=0.75\textwidth]{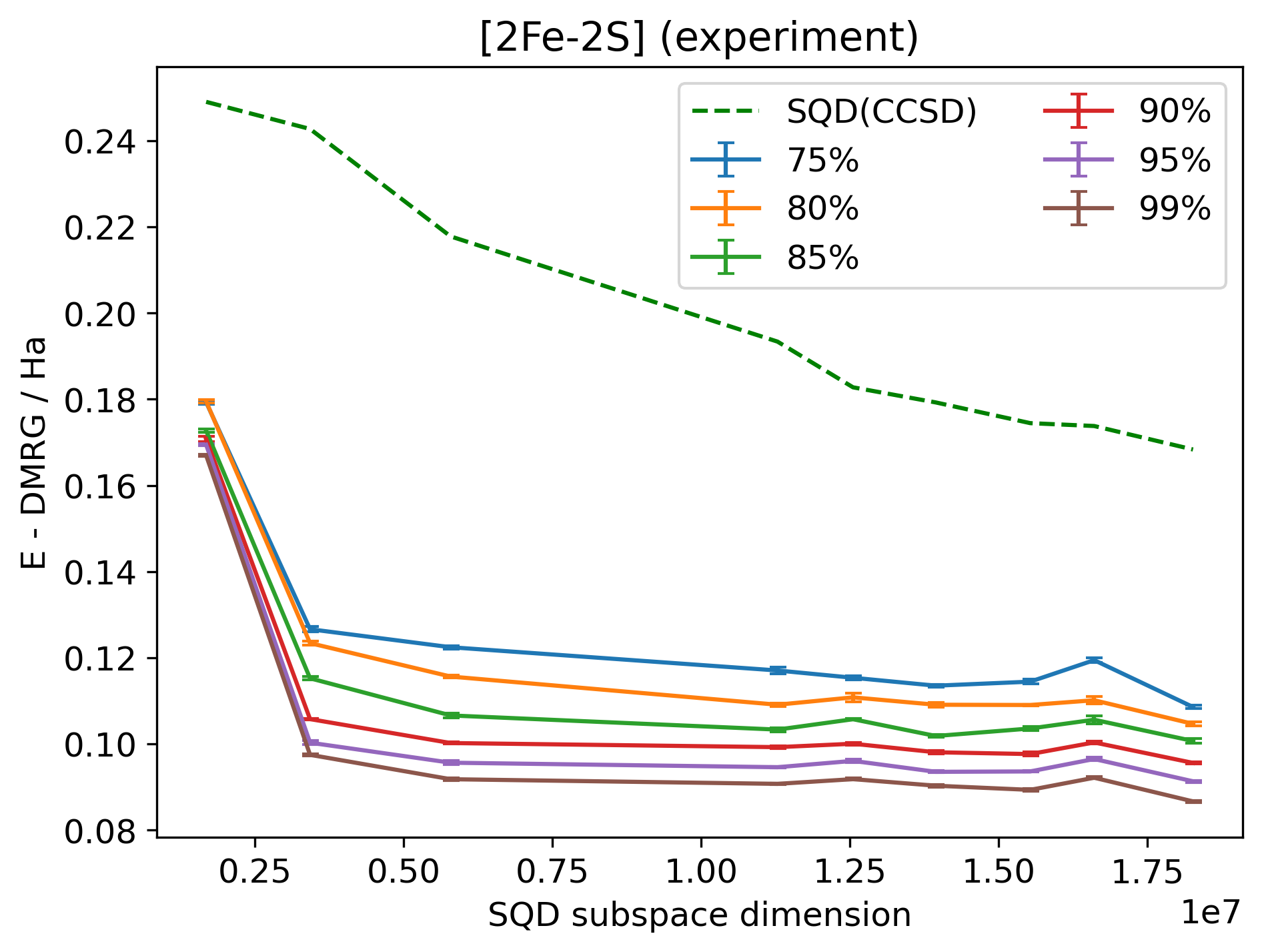}
\caption{Energy errors for the $\fes$ model system from SQD (samples drawn from a single-layer LUCJ wavefunction with parameters from CCSD) and ph-AFQMC-SQD with six truncation levels of CI weight ranging from 75\% to 99\%.}
\label{fig:fe2s2exp}
\end{figure}

SQD and ph-AFQMC-SQD results for the $\fes$ model cluster are shown in Figure \ref{fig:fe2s2exp}. We use as a benchmark energy the density matrix renormalization group (DMRG) value from Ref. \citenum{li_spin-projected_2017} with bond-dimension 2000, which was found to be $\mathcal{O}(1)$ microHartree lower in energy than a DMRG calculation that we performed with bond-dimension 1000; hence, while not explicitly extrapolated, we view this DMRG energy as converged.  The SQD wavefunction was obtained from quantum hardware with LUCJ parameters taken from the CCSD level.  Relative to the SQD energies, ph-AFQMC-SQD energies are much improved, though we find that the amount of improvement to the correlation energy decreases as the SQD subspace dimension is increased (at least in the range of $\scientific{0.25}{7}$ to $\scientific{1.8}{7}$).  In addition, we find that increasing the CI weight percentage retained in the truncated SQD trial wavefunction from 95\% to 98\% does not dramatically change the ph-AFQMC-SQD energies, and also that the ph-AFQMC energy seems to saturate (flat-line) by $\scientific{0.6}{7}$ subspace dimension.  In our view, the latter is an encouraging result, since the major enhancements to the SQD correlation energy, from ph-AFQMC-SQD, are seen already at relatively small subspace dimension; the combination of SQD with ph-AFQMC thus can reduce the burden on the quantum device while retaining relatively high accuracy of the combined method. 

\begin{figure}[H]
\centering
\includegraphics[width=0.75\textwidth]{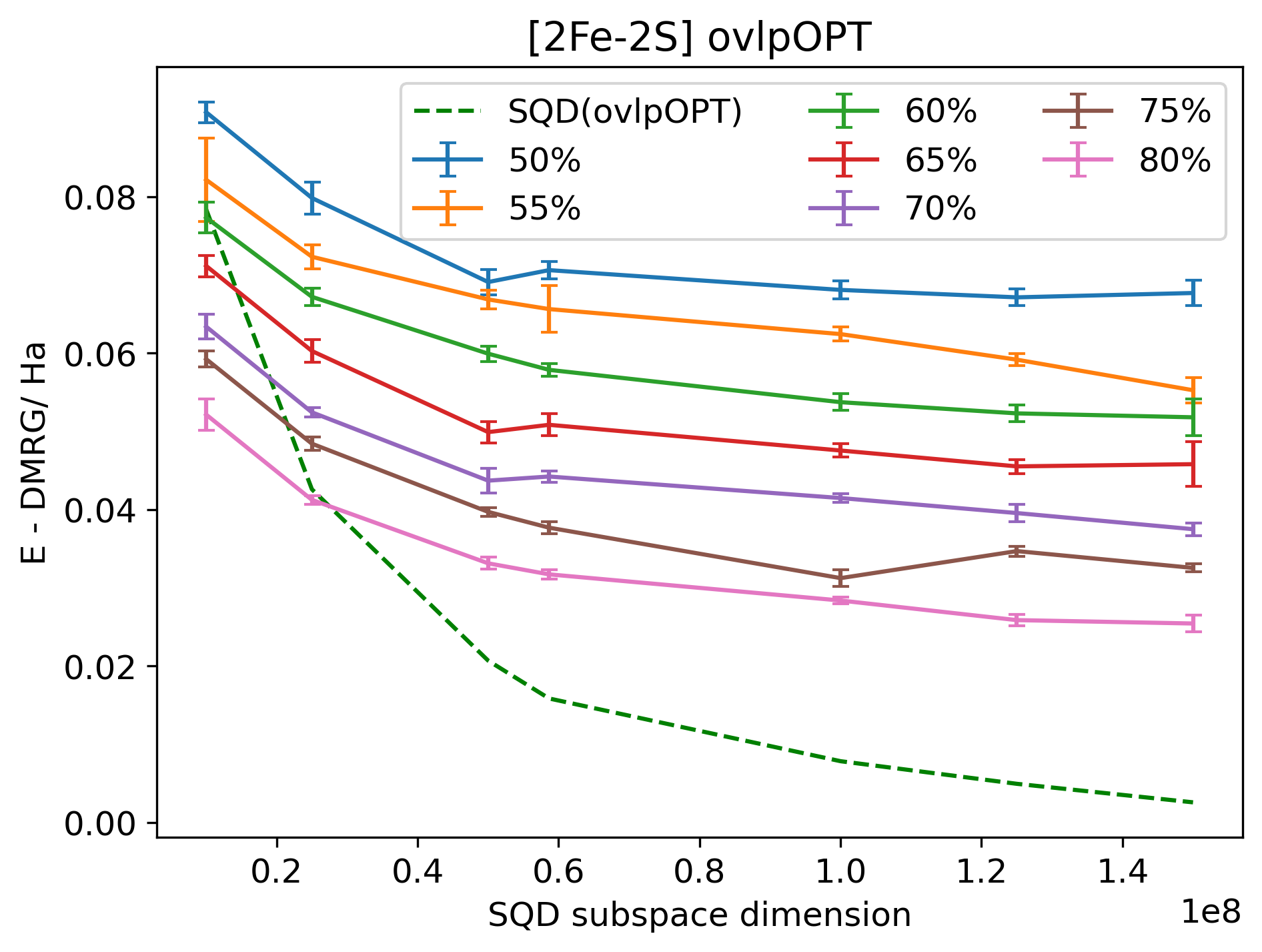}
\caption{Deviations of SQD (dashed green line) and ph-AFQMC-SQD (solid lines) from DMRG energies for the $\fes$ cluster, in Hartree, using SQD trial wavefunctions with increasingly larger subspace dimension (from $10^7$ to $\scientific{1.5}{8}$) and various truncation thresholds of the SQD trial wavefunctions (colored lines from 50 \% to 80 \%).}
\label{fig:fe2s2opt}
\end{figure}

In what follows, we consider the SQD(ovlpOPT) protocol (which relies on prior knowledge of the near-exact wavefunction, as explained in Section 2.1).   In a recent study~\cite{robledo2024chemistry}, this was found to produce very accurate SQD energies and can thus be viewed as a hypothetical example of a future improved SQD protocol.  We consider seven different SQD subspace dimensions, in the range of $\scientific{1.0}{7}$ to $\scientific{1.5}{8}$; for each subspace size, we also use a series of trial wavefunctions for ph-AFQMC with different percentages of retained CI weight, ranging from 50-80\%.   The results are shown in Figure \ref{fig:fe2s2opt}.  As expected, for a fixed CI percentage kept in the ph-AFQMC trial wavefunction, the ph-AFQMC-SQD energy decreases monotonically as the SQD subspace dimension increases.  For a fixed subspace dimension, the ph-AFQMC-SQD energy can also be lowered by increasing the percentage of CI-weight retained in the trial; we stop at 80\% in light of the apparent diminishing marginal energy lowering.  Due to the way that the SQD wavefunctions were obtained in the SQD(ovlpOPT) protocol, the SQD energy without ph-AFQMC post-processing is less than or equal to the ph-AFQMC-SQD results at the subspace size of $\scientific{2.5}{7}$ and beyond.  For smaller SQD subspace dimensions, ph-AFQMC is expected to improve the correlation energy, as was found above in the quantum hardware SQD experiments shown in Figure \ref{fig:fe2s2exp}.

\begin{figure}[H]
\centering
\includegraphics[width=0.75\textwidth]{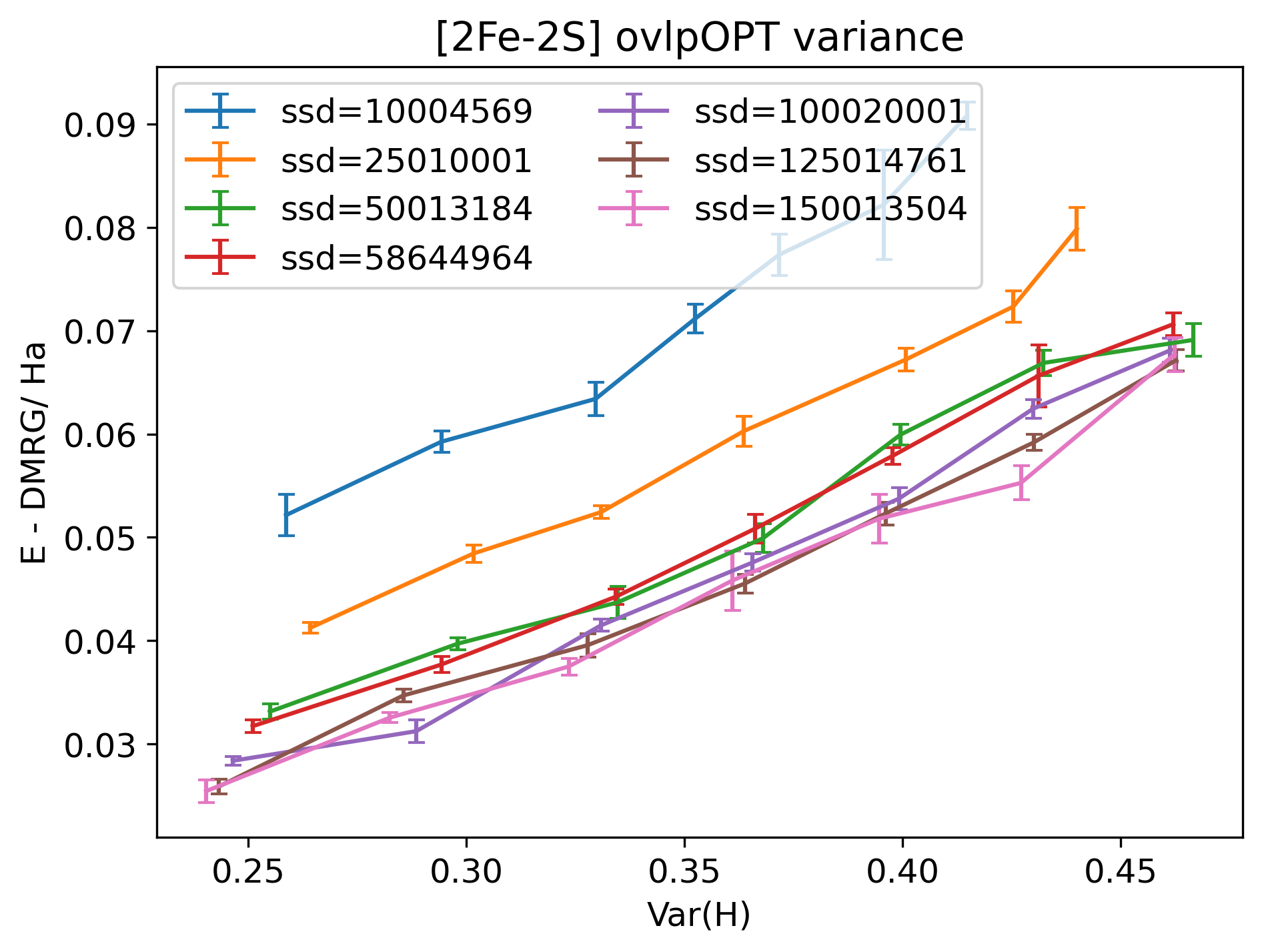}
\caption{Deviations between AFQMC and DMRG energies for the $\fes$ cluster, in Hartree, as a function of the energy variance of the SQD trial wavefunctions, using trials of variable number of configurations (colored lines) and, for each such trial, increasing truncation thresholds (left to right for more conservative to aggressive truncations).}
\label{fig:fe2s2opt2}
\end{figure}

In Figure \ref{fig:fe2s2opt2} we plot ph-AFQMC-SQD energies versus the energy variance of the SQD trial wavefunctions employed.  In the regime of small variances, it is known that a linear relationship exists.  For each subspace dimension, we extrapolate the ph-AFQMC-SQD energies to the limit in which the trial wavefunction has zero-variance; in this limit, ph-AFQMC will be exact, in principle.  In practice, we use a 3-point extrapolation using trials with 70, 75, and 80\% weight.  The extrapolated ph-AFQMC-SQD energies for each subspace size are shown in Table \ref{fe2s2_opt_extrap}.  The energy errors vs the reference DMRG value are much improved by trial-variance extrapolation, in the range of 0.6-13 mHa.  While additional test cases are needed to make any general claims, we find these extrapolated results to be encouraging. %it is encouraging that trial wavefunctions from an SQD procedure on quantum hardware that drew samples from a single-layer (and unoptimized) LUCJ circuit can, with ph-AFQMC postprocessing and variance extrapolation, approach, and in some cases be within, the accuracy target of $\leq$1.6 mHa. 

\begin{table}[htb]
\begin{center}
\begin{tabular}{c|c} 
 SQD subspace dimension & Energy (relative to DMRG) / Ha\\
 \hline
  10004569 & 0.012 \\
  25010001 & -0.003 \\
  50013184 & -0.001 \\
  58644964 & -0.006 \\
  100020001 & -0.011 \\
  125014761 & -0.013 \\
  150013504 & -0.009 \\
\end{tabular}
\caption{Subspace dimension and 3-point extrapolated energies  using ph-AFQMC  with SQD(ovlpOPT) trials for the $\fes$ system.}
\label{fe2s2_opt_extrap}
\end{center}
\end{table}

\subsection{Comparison of SQD and HCI trials for $\fes$}

The rapid pace of development of both quantum hardware capabilities and, e.g., more optimal SQD procedures than SQD(CCSD) make any conclusive, direct comparison of ph-AFQMC-SQD with state-of-the-art classical trial wavefunctions very challenging.  Yet, in this spirit, we present two studies on the $\fes$ model system that shed some light on the relative sampling efficiency of SQD and classical HCI trial wavefunctions themselves (as a function of subspace dimension), and the relative accuracy of the ph-AFQMC energies using these different trials with comparable numbers of determinants kept in the trial wavefunctions.

Fig. \ref{fig:fe2s2opt2} shows that the SQD(CCSD) protocol run previously on quantum hardware has suboptimally sampled  important configurations, relative to the classical HCI procedure.  As expected, the hypothetical SQD(ovlpOPT) protocol produces more accurate energies than HCI with the same number of determinants in the subspace, though both converge to the same energy in the limit of full subspace sampling.   

Within SQD, increasing the number of samples drawn from a quantum device allows diagonalizing the Hamiltonian in higher-dimensional subspaces, roughly $d \leq (2 N_s)^2$ with $N_s$ number of samples and $d$ dimension of the subspace. On the other hand, within VQE, increasing the number of samples drawn from a quantum device allows reducing the statistical uncertainty on the expectation value of the Hamiltonian over the VQE wavefunction, roughly $\sigma \leq \Lambda / N_s^{-1/2}$ where $\Lambda$ is a quantity that depends on the strategy used to measure the Hamiltonian~\cite{wecker2015progress,huggins2021efficient}.
Therefore, comparing the number of samples required by SQD and VQE is very subtle, because samples are used to lower the energy in one case, and to statistically resolve it in the other, with shot and device noise known to induce barren plateaus in the VQE optimization~\cite{wang2021noise}.
This makes reducing the statistical uncertainties on the VQE energy highly desirable. However, due to the scaling with $N_s$ and the magnitude of $\Lambda$, reducing $\sigma$ (e.g. below kcal/mol scale) is generally considered an expensive operation, for example Ref.~\citenum{wecker2015progress} reports $10^{13}$ samples per energy evaluation for $\fes$ in a minimal basis set, whereas our SQD calculations used less than $\scientific{3}{6}$ samples~\cite{robledo2024chemistry}. We note that intense research is currently ongoing on the design of more efficient Hamiltonian measurement strategies~\cite{patel2025quantum}, which may lead to more economical VQE simulations.

%\js{Mario, please add VQE-related comments here} 

\begin{figure}[H]
\centering
\includegraphics[width=0.75\textwidth]{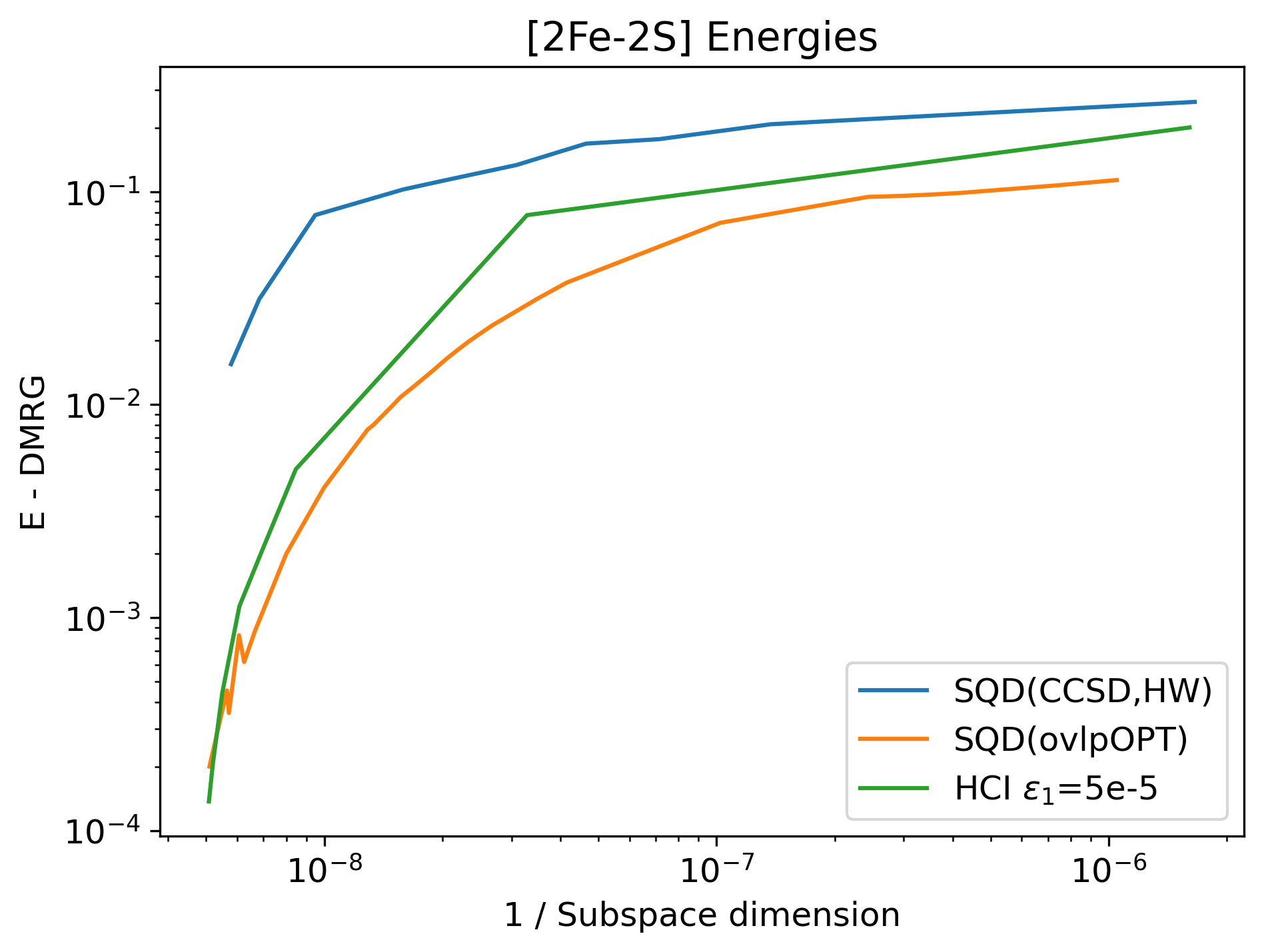}
\caption{Log-log plot of energy error relative to DMRG vs reciprocal subspace dimension, for SQD(CCSD), SQD(ovlpOPT), and classical HCI with a representative value of $\epsilon_1$.}
\label{fig:loglog}
\end{figure}

A comparison of various HCI and SQD wavefunctions, along with the resulting ph-AFQMC energies using truncated versions of these as trial wavefunctions, is shown in Table \ref{table:fe2s2_shcitable}.  The $\epsilon_1$ values investigated are typical of practical classical HCI calculations.  While all of the SQD wavefunctions correspond to a larger subspace dimension than the HCI wavefunctions (at times by an order of magnitude), after truncation to trial wavefunctions of 4597 or 6277 determinants, similar ph-AFQMC energies can be achieved.  
The SQD(ovlpOPT) trial wavefunction, after fairly aggressive truncation to a similar number of determinants, is capable of yielding a ph-AFQMC energy well-below the calculations with HCI and SQD(CCSD) trials. %, though still roughly 70 mHa away from the DMRG reference.
We note that a recent classical ph-AFQMC study with multi-determinant trial wavefunctions obtained using natural orbitals obtained from a near-exact DMRG calculation found that $\mathcal{O}(10^5-10^6)$ determinants were required to get within 1.6 mHa of the exact energy.\cite{huang2024gpu}

\begin{table}[htb]
\begin{center}
\begin{tabular}{c|r|c|r|c} 
 Trial & Subspace dim & CI cutoff (\%) & trunc. N$_{\text{det}}$ & ph-AFQMC E [Ha]\\
 \hline
%  SHCI $\epsilon_1$ = 1x10$^{-4}$   & 854 345& 75 & 2967 & -116.511 5 ± 0.000 9 \\
  HCI $\epsilon_1$ = 1 $\cdot$ 10$^{-4}$   & 854 345& 80 & 4 702 & -116.515 1 ± 0.000 9 \\
  HCI $\epsilon_1$ = 5 $\cdot$ 10$^{-5}$ & 2 041 611& 75 & 3 951 & -116.514 5 ± 0.000 6 \\
  HCI $\epsilon_1$ = 5 $\cdot$ 10$^{-5}$ & 2 041 611& 80 & 6 311 & -116.518 8 ± 0.000 4 \\
   \hline
  %SQD(CCSD)  & 3 426 201 & 90 &  330 & -116.499 845 ± 0.000 185 \\ 
  %SQD(CCSD)  & 3 426 201 & 95 &  949 & -116.505 320 ± 0.000 450 \\ 
  SQD(CCSD)  & 3 426 201 & 99 &  6 277 & -116.508 1 ± 0.000 2 \\ 
  
  %SQD(CCSD)  & 18 267 076 & 90 & 1469 & -116.510 016 ± 0.000 199 \\ 
  SQD(CCSD)  & 18 267 076 & 95 & 4 597 & -116.514 3 ± 0.000 3 \\ 
  %SQD(CCSD)  & 18 267 076 & 99 & 35 535 & -116.519 0 ± 0.000 2 \\ 
  \hline
  SQD(OvlpOPT) &  10 004 569 & 65 & 4 735 & -116.534 4 ± 0.001 4 \\ 
\end{tabular}
\caption{ph-AFQMC results using HCI trials for $\fes$ cluster for two values of configuration selection threshold ($\epsilon_1$) at two levels of truncation.  DMRG reference energy is -116.605 609 Ha.}  
\label{table:fe2s2_shcitable}
\end{center}
\end{table}

\section{Conclusions and Outlook}

%\js{Revise. Mention carry over strategy?}
In this work we explored the use of wavefunctions from sample-based quantum diagonalization (SQD), some of which were obtained from quantum hardware, as trial wavefunctions for phaseless auxiliary-field quantum Monte Carlo (ph-AFQMC).  Our combined algorithm represents a compelling alternative to previously proposed hybrid quantum-classical ph-AFQMC algorithms, primarily because the ph-AFQMC-SQD procedure does not require wavefunction tomography (the VQE prerequisite is a practical bottleneck on near-term quantum hardware).  In addition, our proposed hybrid algorithm involves only a one-time quantum component followed by purely classical ph-AFQMC.  
We have investigated the dissociation of N$_2$ in a 10 electron and 26  orbital space (cc-pVDZ with frozen-core) and an $\fes$ model cluster in a 30 electron and 20 orbital space (Fe 3$d$ and S 3$p$ shells plus four ligand orbitals).  These are challenging systems in light of the available quantum hardware, both in terms of molecular size and the degree of strong correlation in their ground states; our work thus represents a realistic test case of how a quantum-centric algorithm can perform with noisy quantum devices.

We find that ph-AFQMC with limited, but hardware-accessible, SQD trial wavefunctions recovers a significant amount of correlation energy over SQD alone.  The classical ph-AFQMC step, with low-polynomial compute cost scaling with system size,  exploits the embarrassingly parallel nature of QMC and the algorithm's suitability for acceleration on graphical processing units.  Furthermore, we show how extrapolating ph-AFQMC-SQD energies to the zero-trial-variance limit enables the combination of SQD and ph-AFQMC to obtain $\mathcal{O}$(1-10) mHa accuracy vs reference energies for the $\fes$ cluster. 
In practical regimes of limited sample count, combining SQD with ph-AFQMC has the potential to reduce the sampling burden of the former to achieve a fixed correlation energy error.  Indeed, small-weighted configurations can be discarded from the trial wavefunction with relatively small effect on the resulting ph-AFQMC energy.

From the ph-AFQMC development standpoint, in our view the most pressing open problem is understanding the phaseless constraint and its dependence on different trial wavefunction forms.  On purely classical devices, ph-AFQMC with non-linear trial wavefunctions (e.g., of LUCJ form) are, generally, computationally infeasible for all but the smallest system sizes.  One could view the SQD procedure as a way to sample approximate linear wavefunction expansions (amenable to guide ph-AFQMC random walks) using the distribution from the non-linear LUCJ ansatz; in this light, the use of such SQD trial wavefunctions and their effect on the phaseless constraint is of fundamental interest.  However, we emphasize that the optimal wavefunction used to sample SQD configurations  need not be of LUCJ form (in fact, it is certainly not of FCI form).  Future work will seek to address these lines of thought, in addition to exploring possible regimes in which SQD wavefunctions might be more effective or efficient as ph-AFQMC trial wavefunctions than those from purely classical methods such as various flavors of selected CI. 
Such exploration may require sampling configurations from LUCJ circuits with higher depth (we use only a single-layered ansatz here), parametrized quantum circuits other than LUCJ (e.g. other flavors of unitary coupled-cluster), or time-evolution circuits based on sample-based Krylov quantum diagonalization~\cite{mikkelsen2025quantumselectedconfigurationinteractiontimeevolved, sugisaki2025hamiltoniansimulationbasedquantumselectedconfiguration, yu2025quantum} as quantum hardware further matures towards fault-tolerant architectures.

\section{Acknowledgements}
We are grateful to Antonio Mezzacapo for insightful comments on the manuscript.  This work used resources of the Oak Ridge Leadership Computing Facility, which is a DOE Office of Science User Facility supported under Contract DE-AC05-00OR22725.  J. Shee acknowledges support from the Robert A. Welch Foundation, Award No. C-2212.

\noindent At the time of submission, we became aware of Ref. \citenum{yoshida2025auxiliary} by Mizukami and coworkers, which also proposes to use ph-AFQMC with QSCI trial wavefunctions.  We verify that our work was done independently, and without knowledge of that preprint, and view both as valuable and distinct scientific contributions.

\bibliography{main}

\providecommand{\latin}[1]{#1}
\makeatletter
\providecommand{\doi}
  {\begingroup\let\do\@makeother\dospecials
  \catcode`\{=1 \catcode`\}=2 \doi@aux}
\providecommand{\doi@aux}[1]{\endgroup\texttt{#1}}
\makeatother
\providecommand*\mcitethebibliography{\thebibliography}
\csname @ifundefined\endcsname{endmcitethebibliography}  {\let\endmcitethebibliography\endthebibliography}{}
\begin{mcitethebibliography}{91}
\providecommand*\natexlab[1]{#1}
\providecommand*\mciteSetBstSublistMode[1]{}
\providecommand*\mciteSetBstMaxWidthForm[2]{}
\providecommand*\mciteBstWouldAddEndPuncttrue
  {\def\EndOfBibitem{\unskip.}}
\providecommand*\mciteBstWouldAddEndPunctfalse
  {\let\EndOfBibitem\relax}
\providecommand*\mciteSetBstMidEndSepPunct[3]{}
\providecommand*\mciteSetBstSublistLabelBeginEnd[3]{}
\providecommand*\EndOfBibitem{}
\mciteSetBstSublistMode{f}
\mciteSetBstMaxWidthForm{subitem}{(\alph{mcitesubitemcount})}
\mciteSetBstSublistLabelBeginEnd
  {\mcitemaxwidthsubitemform\space}
  {\relax}
  {\relax}

\bibitem[Cao \latin{et~al.}(2019)Cao, Romero, Olson, Degroote, Johnson, Kieferov{\'a}, Kivlichan, Menke, Peropadre, Sawaya, \latin{et~al.} others]{cao2019quantum}
Cao,~Y.; Romero,~J.; Olson,~J.~P.; Degroote,~M.; Johnson,~P.~D.; Kieferov{\'a},~M.; Kivlichan,~I.~D.; Menke,~T.; Peropadre,~B.; Sawaya,~N.~P.; others Quantum chemistry in the age of quantum computing. \emph{Chemical Reviews} \textbf{2019}, \emph{119}, 10856--10915\relax
\mciteBstWouldAddEndPuncttrue
\mciteSetBstMidEndSepPunct{\mcitedefaultmidpunct}
{\mcitedefaultendpunct}{\mcitedefaultseppunct}\relax
\EndOfBibitem
\bibitem[Bauer \latin{et~al.}(2020)Bauer, Bravyi, Motta, and Chan]{bauer2020quantum}
Bauer,~B.; Bravyi,~S.; Motta,~M.; Chan,~G. K.-L. Quantum algorithms for quantum chemistry and quantum materials science. \emph{Chemical Reviews} \textbf{2020}, \emph{120}, 12685--12717\relax
\mciteBstWouldAddEndPuncttrue
\mciteSetBstMidEndSepPunct{\mcitedefaultmidpunct}
{\mcitedefaultendpunct}{\mcitedefaultseppunct}\relax
\EndOfBibitem
\bibitem[McArdle \latin{et~al.}(2020)McArdle, Endo, Aspuru-Guzik, Benjamin, and Yuan]{mcardle2020quantum}
McArdle,~S.; Endo,~S.; Aspuru-Guzik,~A.; Benjamin,~S.~C.; Yuan,~X. Quantum computational chemistry. \emph{Reviews of Modern Physics} \textbf{2020}, \emph{92}, 015003\relax
\mciteBstWouldAddEndPuncttrue
\mciteSetBstMidEndSepPunct{\mcitedefaultmidpunct}
{\mcitedefaultendpunct}{\mcitedefaultseppunct}\relax
\EndOfBibitem
\bibitem[Cerezo \latin{et~al.}(2021)Cerezo, Arrasmith, Babbush, Benjamin, Endo, Fujii, McClean, Mitarai, Yuan, Cincio, \latin{et~al.} others]{cerezo2021variational}
Cerezo,~M.; Arrasmith,~A.; Babbush,~R.; Benjamin,~S.~C.; Endo,~S.; Fujii,~K.; McClean,~J.~R.; Mitarai,~K.; Yuan,~X.; Cincio,~L.; others Variational quantum algorithms. \emph{Nature Reviews Physics} \textbf{2021}, \emph{3}, 625--644\relax
\mciteBstWouldAddEndPuncttrue
\mciteSetBstMidEndSepPunct{\mcitedefaultmidpunct}
{\mcitedefaultendpunct}{\mcitedefaultseppunct}\relax
\EndOfBibitem
\bibitem[Motta and Rice(2022)Motta, and Rice]{motta2022emerging}
Motta,~M.; Rice,~J.~E. Emerging quantum computing algorithms for quantum chemistry. \emph{Wiley Interdisciplinary Reviews: Computational Molecular Science} \textbf{2022}, \emph{12}, e1580\relax
\mciteBstWouldAddEndPuncttrue
\mciteSetBstMidEndSepPunct{\mcitedefaultmidpunct}
{\mcitedefaultendpunct}{\mcitedefaultseppunct}\relax
\EndOfBibitem
\bibitem[McClean \latin{et~al.}(2016)McClean, Romero, Babbush, and Aspuru-Guzik]{mcclean2016theory}
McClean,~J.~R.; Romero,~J.; Babbush,~R.; Aspuru-Guzik,~A. The theory of variational hybrid quantum-classical algorithms. \emph{New Journal of Physics} \textbf{2016}, \emph{18}, 023023\relax
\mciteBstWouldAddEndPuncttrue
\mciteSetBstMidEndSepPunct{\mcitedefaultmidpunct}
{\mcitedefaultendpunct}{\mcitedefaultseppunct}\relax
\EndOfBibitem
\bibitem[Alexeev \latin{et~al.}(2024)Alexeev, Amsler, Barroca, Bassini, Battelle, Camps, Casanova, Choi, Chong, Chung, \latin{et~al.} others]{alexeev2024quantum}
Alexeev,~Y.; Amsler,~M.; Barroca,~M.~A.; Bassini,~S.; Battelle,~T.; Camps,~D.; Casanova,~D.; Choi,~Y.~J.; Chong,~F.~T.; Chung,~C.; others Quantum-centric supercomputing for materials science: A perspective on challenges and future directions. \emph{Future Generation Computer Systems} \textbf{2024}, \emph{160}, 666--710\relax
\mciteBstWouldAddEndPuncttrue
\mciteSetBstMidEndSepPunct{\mcitedefaultmidpunct}
{\mcitedefaultendpunct}{\mcitedefaultseppunct}\relax
\EndOfBibitem
\bibitem[Huang \latin{et~al.}(2023)Huang, Li, and Evangelista]{huang2023leveraging}
Huang,~R.; Li,~C.; Evangelista,~F.~A. Leveraging small-scale quantum computers with unitarily downfolded hamiltonians. \emph{PRX Quantum} \textbf{2023}, \emph{4}, 020313\relax
\mciteBstWouldAddEndPuncttrue
\mciteSetBstMidEndSepPunct{\mcitedefaultmidpunct}
{\mcitedefaultendpunct}{\mcitedefaultseppunct}\relax
\EndOfBibitem
\bibitem[Fitzpatrick \latin{et~al.}(2024)Fitzpatrick, Talarico, Eik{\aa}s, and Knecht]{fitzpatrick2024quantum}
Fitzpatrick,~A.; Talarico,~N.~W.; Eik{\aa}s,~R. D.~R.; Knecht,~S. Quantum-centric strong and dynamical electron correlation: A resource-efficient second-order $ N $-electron valence perturbation theory formulation for near-term quantum devices. \emph{arXiv preprint arXiv:2405.15422} \textbf{2024}, \relax
\mciteBstWouldAddEndPunctfalse
\mciteSetBstMidEndSepPunct{\mcitedefaultmidpunct}
{}{\mcitedefaultseppunct}\relax
\EndOfBibitem
\bibitem[Motta \latin{et~al.}(2020)Motta, Gujarati, Rice, Kumar, Masteran, Latone, Lee, Valeev, and Takeshita]{motta2020quantum}
Motta,~M.; Gujarati,~T.~P.; Rice,~J.~E.; Kumar,~A.; Masteran,~C.; Latone,~J.~A.; Lee,~E.; Valeev,~E.~F.; Takeshita,~T.~Y. Quantum simulation of electronic structure with a transcorrelated Hamiltonian: improved accuracy with a smaller footprint on the quantum computer. \emph{Physical Chemistry Chemical Physics} \textbf{2020}, \emph{22}, 24270--24281\relax
\mciteBstWouldAddEndPuncttrue
\mciteSetBstMidEndSepPunct{\mcitedefaultmidpunct}
{\mcitedefaultendpunct}{\mcitedefaultseppunct}\relax
\EndOfBibitem
\bibitem[Erhart \latin{et~al.}(2024)Erhart, Yoshida, Khinevich, and Mizukami]{erhart2024coupled}
Erhart,~L.; Yoshida,~Y.; Khinevich,~V.; Mizukami,~W. Coupled cluster method tailored with quantum computing. \emph{Physical Review Research} \textbf{2024}, \emph{6}, 023230\relax
\mciteBstWouldAddEndPuncttrue
\mciteSetBstMidEndSepPunct{\mcitedefaultmidpunct}
{\mcitedefaultendpunct}{\mcitedefaultseppunct}\relax
\EndOfBibitem
\bibitem[Tammaro \latin{et~al.}(2023)Tammaro, Galli, Rice, and Motta]{tammaro2023n}
Tammaro,~A.; Galli,~D.~E.; Rice,~J.~E.; Motta,~M. N-Electron valence perturbation theory with reference wave functions from quantum computing: Application to the relative stability of hydroxide anion and hydroxyl radical. \emph{The Journal of Physical Chemistry A} \textbf{2023}, \emph{127}, 817--827\relax
\mciteBstWouldAddEndPuncttrue
\mciteSetBstMidEndSepPunct{\mcitedefaultmidpunct}
{\mcitedefaultendpunct}{\mcitedefaultseppunct}\relax
\EndOfBibitem
\bibitem[Khinevich and Mizukami(2025)Khinevich, and Mizukami]{khinevich2025enhancing}
Khinevich,~V.; Mizukami,~W. Enhancing quantum computations with the synergy of auxiliary field quantum Monte Carlo and computational basis tomography. \emph{arXiv preprint arXiv:2502.20066} \textbf{2025}, \relax
\mciteBstWouldAddEndPunctfalse
\mciteSetBstMidEndSepPunct{\mcitedefaultmidpunct}
{}{\mcitedefaultseppunct}\relax
\EndOfBibitem
\bibitem[Huggins \latin{et~al.}(2022)Huggins, O’Gorman, Rubin, Reichman, Babbush, and Lee]{huggins2022unbiasing}
Huggins,~W.~J.; O’Gorman,~B.~A.; Rubin,~N.~C.; Reichman,~D.~R.; Babbush,~R.; Lee,~J. Unbiasing fermionic quantum {M}onte {C}arlo with a quantum computer. \emph{Nature} \textbf{2022}, \emph{603}, 416--420\relax
\mciteBstWouldAddEndPuncttrue
\mciteSetBstMidEndSepPunct{\mcitedefaultmidpunct}
{\mcitedefaultendpunct}{\mcitedefaultseppunct}\relax
\EndOfBibitem
\bibitem[Zhang and Krakauer(2003)Zhang, and Krakauer]{zhang2003quantum}
Zhang,~S.; Krakauer,~H. Quantum {M}onte {C}arlo method using phase-free random walks with {S}later determinants. \emph{Physical Review Letters} \textbf{2003}, \emph{90}, 136401\relax
\mciteBstWouldAddEndPuncttrue
\mciteSetBstMidEndSepPunct{\mcitedefaultmidpunct}
{\mcitedefaultendpunct}{\mcitedefaultseppunct}\relax
\EndOfBibitem
\bibitem[Stratonovich(1957)]{stratonovich_method_1957}
Stratonovich,~R.~L. On a method of calculating quantum distribution functions. \emph{Soviet Physics Doklady} \textbf{1957}, \emph{2}, 416, ADS Bibcode: 1957SPhD....2..416S\relax
\mciteBstWouldAddEndPuncttrue
\mciteSetBstMidEndSepPunct{\mcitedefaultmidpunct}
{\mcitedefaultendpunct}{\mcitedefaultseppunct}\relax
\EndOfBibitem
\bibitem[Hubbard(1959)]{hubbard_calculation_1959}
Hubbard,~J. Calculation of {Partition} {Functions}. \emph{Physical Review Letters} \textbf{1959}, \emph{3}, 77--78\relax
\mciteBstWouldAddEndPuncttrue
\mciteSetBstMidEndSepPunct{\mcitedefaultmidpunct}
{\mcitedefaultendpunct}{\mcitedefaultseppunct}\relax
\EndOfBibitem
\bibitem[Peruzzo \latin{et~al.}(2014)Peruzzo, McClean, Shadbolt, Yung, Zhou, Love, Aspuru-Guzik, and O’brien]{peruzzo2014variational}
Peruzzo,~A.; McClean,~J.; Shadbolt,~P.; Yung,~M.-H.; Zhou,~X.-Q.; Love,~P.~J.; Aspuru-Guzik,~A.; O’brien,~J.~L. A variational eigenvalue solver on a photonic quantum processor. \emph{Nature communications} \textbf{2014}, \emph{5}, 4213\relax
\mciteBstWouldAddEndPuncttrue
\mciteSetBstMidEndSepPunct{\mcitedefaultmidpunct}
{\mcitedefaultendpunct}{\mcitedefaultseppunct}\relax
\EndOfBibitem
\bibitem[Aaronson(2018)]{aaronson2018shadow}
Aaronson,~S. Shadow tomography of quantum states. Proceedings of the 50th annual ACM SIGACT symposium on theory of computing. 2018; pp 325--338\relax
\mciteBstWouldAddEndPuncttrue
\mciteSetBstMidEndSepPunct{\mcitedefaultmidpunct}
{\mcitedefaultendpunct}{\mcitedefaultseppunct}\relax
\EndOfBibitem
\bibitem[Zhao \latin{et~al.}(2021)Zhao, Rubin, and Miyake]{zhao2021fermionic}
Zhao,~A.; Rubin,~N.~C.; Miyake,~A. Fermionic partial tomography via classical shadows. \emph{Physical Review Letters} \textbf{2021}, \emph{127}, 110504\relax
\mciteBstWouldAddEndPuncttrue
\mciteSetBstMidEndSepPunct{\mcitedefaultmidpunct}
{\mcitedefaultendpunct}{\mcitedefaultseppunct}\relax
\EndOfBibitem
\bibitem[Huang \latin{et~al.}(2024)Huang, Chen, Gupt, Suchara, Tran, McArdle, and Galli]{huang2024evaluating}
Huang,~B.; Chen,~Y.-T.; Gupt,~B.; Suchara,~M.; Tran,~A.; McArdle,~S.; Galli,~G. Evaluating a quantum-classical quantum Monte Carlo algorithm with matchgate shadows. \emph{Physical Review Research} \textbf{2024}, \emph{6}, 043063\relax
\mciteBstWouldAddEndPuncttrue
\mciteSetBstMidEndSepPunct{\mcitedefaultmidpunct}
{\mcitedefaultendpunct}{\mcitedefaultseppunct}\relax
\EndOfBibitem
\bibitem[Zhao \latin{et~al.}(2025)Zhao, Goings, Aboumrad, Arrasmith, Calderin, Churchill, Gabay, Harvey-Brown, Hiles, Kaja, Keesan, Kulesz, Maksymov, Maruo, Muñoz, Nijholt, Schiller, de~Sereville, Smidutz, Tripier, Yao, Zaveri, Collins, Roetteler, Epifanovsky, Kovyrshin, Tornberg, Broo, Hammond, Chandani, Khalate, Kyoseva, Chen, Kessler, Lin, Ramu, Shaffer, Brett, Huang, Hugues, and Takeshita]{zhao2025quantumclassicalauxiliaryfieldquantum}
Zhao,~L.; Goings,~J.~J.; Aboumrad,~W.; Arrasmith,~A.; Calderin,~L.; Churchill,~S.; Gabay,~D.; Harvey-Brown,~T.; Hiles,~M.; Kaja,~M.; Keesan,~M.; Kulesz,~K.; Maksymov,~A.; Maruo,~M.; Muñoz,~M.; Nijholt,~B.; Schiller,~R.; de~Sereville,~Y.; Smidutz,~A.; Tripier,~F.; Yao,~G.; Zaveri,~T.; Collins,~C.; Roetteler,~M.; Epifanovsky,~E.; Kovyrshin,~A.; Tornberg,~L.; Broo,~A.; Hammond,~J.~R.; Chandani,~Z.; Khalate,~P.; Kyoseva,~E.; Chen,~Y.-T.; Kessler,~E.~M.; Lin,~C. Y.-Y.; Ramu,~G.; Shaffer,~R.; Brett,~M.; Huang,~B.; Hugues,~M.~R.; Takeshita,~T.~Y. Quantum-Classical Auxiliary Field Quantum Monte Carlo with Matchgate Shadows on Trapped Ion Quantum Computers. 2025; \url{https://arxiv.org/abs/2506.22408}\relax
\mciteBstWouldAddEndPuncttrue
\mciteSetBstMidEndSepPunct{\mcitedefaultmidpunct}
{\mcitedefaultendpunct}{\mcitedefaultseppunct}\relax
\EndOfBibitem
\bibitem[Kiser \latin{et~al.}(2024)Kiser, Beuerle, and Simkovic~IV]{kiser2024contextual}
Kiser,~M.; Beuerle,~M.; Simkovic~IV,~F. Contextual subspace auxiliary-field quantum Monte Carlo: Improved bias with reduced quantum resources. \emph{arXiv preprint arXiv:2408.06160} \textbf{2024}, \relax
\mciteBstWouldAddEndPunctfalse
\mciteSetBstMidEndSepPunct{\mcitedefaultmidpunct}
{}{\mcitedefaultseppunct}\relax
\EndOfBibitem
\bibitem[Amsler \latin{et~al.}(2023)Amsler, Deglmann, Degroote, Kaicher, Kiser, K{\"u}hn, Kumar, Maier, Samsonidze, Schroeder, \latin{et~al.} others]{amsler2023classical}
Amsler,~M.; Deglmann,~P.; Degroote,~M.; Kaicher,~M.~P.; Kiser,~M.; K{\"u}hn,~M.; Kumar,~C.; Maier,~A.; Samsonidze,~G.; Schroeder,~A.; others Classical and quantum trial wave functions in auxiliary-field quantum Monte Carlo applied to oxygen allotropes and a CuBr$_2$ model system. \emph{The Journal of Chemical Physics} \textbf{2023}, \emph{159}\relax
\mciteBstWouldAddEndPuncttrue
\mciteSetBstMidEndSepPunct{\mcitedefaultmidpunct}
{\mcitedefaultendpunct}{\mcitedefaultseppunct}\relax
\EndOfBibitem
\bibitem[Kiser \latin{et~al.}(2024)Kiser, Schroeder, Anselmetti, Kumar, Moll, Streif, and Vodola]{kiser2024classical}
Kiser,~M.; Schroeder,~A.; Anselmetti,~G.-L.~R.; Kumar,~C.; Moll,~N.; Streif,~M.; Vodola,~D. Classical and quantum cost of measurement strategies for quantum-enhanced auxiliary field quantum Monte Carlo. \emph{New Journal of Physics} \textbf{2024}, \emph{26}, 033022\relax
\mciteBstWouldAddEndPuncttrue
\mciteSetBstMidEndSepPunct{\mcitedefaultmidpunct}
{\mcitedefaultendpunct}{\mcitedefaultseppunct}\relax
\EndOfBibitem
\bibitem[Mazzola and Carleo(2022)Mazzola, and Carleo]{mazzola2022exponential}
Mazzola,~G.; Carleo,~G. Exponential challenges in unbiasing quantum Monte Carlo algorithms with quantum computers. \emph{arXiv preprint arXiv:2205.09203} \textbf{2022}, \relax
\mciteBstWouldAddEndPunctfalse
\mciteSetBstMidEndSepPunct{\mcitedefaultmidpunct}
{}{\mcitedefaultseppunct}\relax
\EndOfBibitem
\bibitem[Kanno \latin{et~al.}(2023)Kanno, Kohda, Imai, Koh, Mitarai, Mizukami, and Nakagawa]{kanno2023quantum}
Kanno,~K.; Kohda,~M.; Imai,~R.; Koh,~S.; Mitarai,~K.; Mizukami,~W.; Nakagawa,~Y.~O. Quantum-selected configuration interaction: Classical diagonalization of Hamiltonians in subspaces selected by quantum computers. \emph{arXiv preprint arXiv:2302.11320} \textbf{2023}, \relax
\mciteBstWouldAddEndPunctfalse
\mciteSetBstMidEndSepPunct{\mcitedefaultmidpunct}
{}{\mcitedefaultseppunct}\relax
\EndOfBibitem
\bibitem[Robledo-Moreno \latin{et~al.}(2025)Robledo-Moreno, Motta, Haas, Javadi-Abhari, Jurcevic, Kirby, Martiel, Sharma, Sharma, Shirakawa, Sitdikov, Sun, Sung, Takita, Tran, Yunoki, and Mezzacapo]{robledo2024chemistry}
Robledo-Moreno,~J.; Motta,~M.; Haas,~H.; Javadi-Abhari,~A.; Jurcevic,~P.; Kirby,~W.; Martiel,~S.; Sharma,~K.; Sharma,~S.; Shirakawa,~T.; Sitdikov,~I.; Sun,~R.-Y.; Sung,~K.~J.; Takita,~M.; Tran,~M.~C.; Yunoki,~S.; Mezzacapo,~A. Chemistry beyond the scale of exact diagonalization on a quantum-centric supercomputer. \emph{Science Advances} \textbf{2025}, \emph{11}, eadu9991\relax
\mciteBstWouldAddEndPuncttrue
\mciteSetBstMidEndSepPunct{\mcitedefaultmidpunct}
{\mcitedefaultendpunct}{\mcitedefaultseppunct}\relax
\EndOfBibitem
\bibitem[Barison \latin{et~al.}(2025)Barison, Robledo~Moreno, and Motta]{barison2024quantum}
Barison,~S.; Robledo~Moreno,~J.; Motta,~M. Quantum-centric computation of molecular excited states with extended sample-based quantum diagonalization. \emph{Quantum Science and Technology} \textbf{2025}, \emph{10}, 025034\relax
\mciteBstWouldAddEndPuncttrue
\mciteSetBstMidEndSepPunct{\mcitedefaultmidpunct}
{\mcitedefaultendpunct}{\mcitedefaultseppunct}\relax
\EndOfBibitem
\bibitem[Kaliakin \latin{et~al.}(2024)Kaliakin, Shajan, Moreno, Li, Mitra, Motta, Johnson, Saki, Das, Sitdikov, \latin{et~al.} others]{kaliakin2024accurate}
Kaliakin,~D.; Shajan,~A.; Moreno,~J.~R.; Li,~Z.; Mitra,~A.; Motta,~M.; Johnson,~C.; Saki,~A.~A.; Das,~S.; Sitdikov,~I.; others Accurate quantum-centric simulations of supramolecular interactions. \emph{arXiv preprint arXiv:2410.09209} \textbf{2024}, \relax
\mciteBstWouldAddEndPunctfalse
\mciteSetBstMidEndSepPunct{\mcitedefaultmidpunct}
{}{\mcitedefaultseppunct}\relax
\EndOfBibitem
\bibitem[Shajan \latin{et~al.}(2024)Shajan, Kaliakin, Mitra, Moreno, Li, Motta, Johnson, Saki, Das, Sitdikov, \latin{et~al.} others]{shajan2024towards}
Shajan,~A.; Kaliakin,~D.; Mitra,~A.; Moreno,~J.~R.; Li,~Z.; Motta,~M.; Johnson,~C.; Saki,~A.~A.; Das,~S.; Sitdikov,~I.; others Towards quantum-centric simulations of extended molecules: sample-based quantum diagonalization enhanced with density matrix embedding theory. \emph{arXiv preprint arXiv:2411.09861} \textbf{2024}, \relax
\mciteBstWouldAddEndPunctfalse
\mciteSetBstMidEndSepPunct{\mcitedefaultmidpunct}
{}{\mcitedefaultseppunct}\relax
\EndOfBibitem
\bibitem[Kaliakin \latin{et~al.}(2025)Kaliakin, Shajan, Liang, and Merz~Jr]{kaliakin2025implicit}
Kaliakin,~D.; Shajan,~A.; Liang,~F.; Merz~Jr,~K.~M. Implicit solvent sample-based quantum diagonalization. \emph{arXiv preprint arXiv:2502.10189} \textbf{2025}, \relax
\mciteBstWouldAddEndPunctfalse
\mciteSetBstMidEndSepPunct{\mcitedefaultmidpunct}
{}{\mcitedefaultseppunct}\relax
\EndOfBibitem
\bibitem[Reinholdt \latin{et~al.}(2025)Reinholdt, Ziems, Kjellgren, Coriani, Sauer, and Kongsted]{reinholdt2025exposing}
Reinholdt,~P.; Ziems,~K.~M.; Kjellgren,~E.~R.; Coriani,~S.; Sauer,~S.; Kongsted,~J. Exposing a fatal flaw in sample-based quantum diagonalization methods. \emph{arXiv preprint arXiv:2501.07231} \textbf{2025}, \relax
\mciteBstWouldAddEndPunctfalse
\mciteSetBstMidEndSepPunct{\mcitedefaultmidpunct}
{}{\mcitedefaultseppunct}\relax
\EndOfBibitem
\bibitem[Wecker \latin{et~al.}(2015)Wecker, Hastings, and Troyer]{wecker2015progress}
Wecker,~D.; Hastings,~M.~B.; Troyer,~M. Progress towards practical quantum variational algorithms. \emph{Physical Review A} \textbf{2015}, \emph{92}, 042303\relax
\mciteBstWouldAddEndPuncttrue
\mciteSetBstMidEndSepPunct{\mcitedefaultmidpunct}
{\mcitedefaultendpunct}{\mcitedefaultseppunct}\relax
\EndOfBibitem
\bibitem[Motta \latin{et~al.}(2023)Motta, Sung, Whaley, Head-Gordon, and Shee]{motta2023bridging}
Motta,~M.; Sung,~K.~J.; Whaley,~K.~B.; Head-Gordon,~M.; Shee,~J. Bridging physical intuition and hardware efficiency for correlated electronic states: the local unitary cluster Jastrow ansatz for electronic structure. \textbf{2023}, \relax
\mciteBstWouldAddEndPunctfalse
\mciteSetBstMidEndSepPunct{\mcitedefaultmidpunct}
{}{\mcitedefaultseppunct}\relax
\EndOfBibitem
\bibitem[Motta \latin{et~al.}(2024)Motta, Sung, and Shee]{motta2024quantum}
Motta,~M.; Sung,~K.~J.; Shee,~J. Quantum algorithms for the variational optimization of correlated electronic states with stochastic reconfiguration and the linear method. \emph{The Journal of Physical Chemistry A} \textbf{2024}, \emph{128}, 8762--8776\relax
\mciteBstWouldAddEndPuncttrue
\mciteSetBstMidEndSepPunct{\mcitedefaultmidpunct}
{\mcitedefaultendpunct}{\mcitedefaultseppunct}\relax
\EndOfBibitem
\bibitem[Matsuzawa and Kurashige(2020)Matsuzawa, and Kurashige]{matsuzawa2020jastrow}
Matsuzawa,~Y.; Kurashige,~Y. Jastrow-type decomposition in quantum chemistry for low-depth quantum circuits. \emph{Journal of Chemical Theory and Computation} \textbf{2020}, \emph{16}, 944--952\relax
\mciteBstWouldAddEndPuncttrue
\mciteSetBstMidEndSepPunct{\mcitedefaultmidpunct}
{\mcitedefaultendpunct}{\mcitedefaultseppunct}\relax
\EndOfBibitem
\bibitem[Ragonneau(2022)]{rago_thesis}
Ragonneau,~T.~M. Model-Based Derivative-Free Optimization Methods and Software. Ph.D.\ thesis, Department of Applied Mathematics, The Hong Kong Polytechnic University, Hong Kong, China, 2022\relax
\mciteBstWouldAddEndPuncttrue
\mciteSetBstMidEndSepPunct{\mcitedefaultmidpunct}
{\mcitedefaultendpunct}{\mcitedefaultseppunct}\relax
\EndOfBibitem
\bibitem[Ragonneau and Zhang(2024)Ragonneau, and Zhang]{razh_cobyqa}
Ragonneau,~T.~M.; Zhang,~Z. {COBYQA} {V}ersion 1.1.2. 2024; \url{https://www.cobyqa.com}\relax
\mciteBstWouldAddEndPuncttrue
\mciteSetBstMidEndSepPunct{\mcitedefaultmidpunct}
{\mcitedefaultendpunct}{\mcitedefaultseppunct}\relax
\EndOfBibitem
\bibitem[Storn and Price(1997)Storn, and Price]{storn1997differential}
Storn,~R.; Price,~K. Differential Evolution -- a simple and efficient heuristic for global optimization over continuous spaces. \emph{Journal of Global Optimization} \textbf{1997}, \emph{11}, 341--359\relax
\mciteBstWouldAddEndPuncttrue
\mciteSetBstMidEndSepPunct{\mcitedefaultmidpunct}
{\mcitedefaultendpunct}{\mcitedefaultseppunct}\relax
\EndOfBibitem
\bibitem[Motta and Zhang(2018)Motta, and Zhang]{motta_ab_2018}
Motta,~M.; Zhang,~S. Ab initio computations of molecular systems by the auxiliary‐field quantum {Monte} {Carlo} method. \emph{Wiley Interdisciplinary Reviews: Computational Molecular Science} \textbf{2018}, \emph{8}, e1364\relax
\mciteBstWouldAddEndPuncttrue
\mciteSetBstMidEndSepPunct{\mcitedefaultmidpunct}
{\mcitedefaultendpunct}{\mcitedefaultseppunct}\relax
\EndOfBibitem
\bibitem[Lee \latin{et~al.}(2022)Lee, Pham, and Reichman]{lee_twenty_2022}
Lee,~J.; Pham,~H.~Q.; Reichman,~D.~R. Twenty years of auxiliary-field quantum Monte Carlo in quantum chemistry: an overview and assessment on main group chemistry and bond-breaking. \emph{Journal of Chemical Theory and Computation} \textbf{2022}, \emph{18}, 7024--7042\relax
\mciteBstWouldAddEndPuncttrue
\mciteSetBstMidEndSepPunct{\mcitedefaultmidpunct}
{\mcitedefaultendpunct}{\mcitedefaultseppunct}\relax
\EndOfBibitem
\bibitem[Shee \latin{et~al.}(2023)Shee, Weber, Reichman, Friesner, and Zhang]{shee_potentially_2023}
Shee,~J.; Weber,~J.~L.; Reichman,~D.~R.; Friesner,~R.~A.; Zhang,~S. On the potentially transformative role of auxiliary-field quantum {Monte} {Carlo} in quantum chemistry: {A} highly accurate method for transition metals and beyond. \emph{The Journal of Chemical Physics} \textbf{2023}, \emph{158}, 140901\relax
\mciteBstWouldAddEndPuncttrue
\mciteSetBstMidEndSepPunct{\mcitedefaultmidpunct}
{\mcitedefaultendpunct}{\mcitedefaultseppunct}\relax
\EndOfBibitem
\bibitem[Shee \latin{et~al.}(2017)Shee, Zhang, Reichman, and Friesner]{shee2017chemical}
Shee,~J.; Zhang,~S.; Reichman,~D.~R.; Friesner,~R.~A. Chemical transformations approaching chemical accuracy via correlated sampling in auxiliary-field quantum {M}onte {C}arlo. \emph{Journal of Chemical Theory and Computation} \textbf{2017}, \emph{13}, 2667--2680\relax
\mciteBstWouldAddEndPuncttrue
\mciteSetBstMidEndSepPunct{\mcitedefaultmidpunct}
{\mcitedefaultendpunct}{\mcitedefaultseppunct}\relax
\EndOfBibitem
\bibitem[Chen \latin{et~al.}(2023)Chen, Yang, Morales, and Zhang]{chen2023algorithm}
Chen,~S.; Yang,~Y.; Morales,~M.; Zhang,~S. Algorithm for branching and population control in correlated sampling. \emph{Physical Review Research} \textbf{2023}, \emph{5}, 043169\relax
\mciteBstWouldAddEndPuncttrue
\mciteSetBstMidEndSepPunct{\mcitedefaultmidpunct}
{\mcitedefaultendpunct}{\mcitedefaultseppunct}\relax
\EndOfBibitem
\bibitem[Weber \latin{et~al.}(2022)Weber, Vuong, Devlaminck, Shee, Lee, Reichman, and Friesner]{weber2022localized}
Weber,~J.~L.; Vuong,~H.; Devlaminck,~P.~A.; Shee,~J.; Lee,~J.; Reichman,~D.~R.; Friesner,~R.~A. A localized-orbital energy evaluation for auxiliary-field quantum {M}onte {C}arlo. \emph{Journal of Chemical Theory and Computation} \textbf{2022}, \relax
\mciteBstWouldAddEndPunctfalse
\mciteSetBstMidEndSepPunct{\mcitedefaultmidpunct}
{}{\mcitedefaultseppunct}\relax
\EndOfBibitem
\bibitem[Kurian \latin{et~al.}(2023)Kurian, Ye, Mahajan, Berkelbach, and Sharma]{kurian2023toward}
Kurian,~J.~S.; Ye,~H.-Z.; Mahajan,~A.; Berkelbach,~T.~C.; Sharma,~S. Toward linear scaling auxiliary-field quantum {M}onte {C}arlo with local natural orbitals. \emph{Journal of Chemical Theory and Computation} \textbf{2023}, \emph{20}, 134--142\relax
\mciteBstWouldAddEndPuncttrue
\mciteSetBstMidEndSepPunct{\mcitedefaultmidpunct}
{\mcitedefaultendpunct}{\mcitedefaultseppunct}\relax
\EndOfBibitem
\bibitem[Shee \latin{et~al.}(2018)Shee, Arthur, Zhang, Reichman, and Friesner]{shee2018phaseless}
Shee,~J.; Arthur,~E.~J.; Zhang,~S.; Reichman,~D.~R.; Friesner,~R.~A. Phaseless auxiliary-field quantum {M}onte {C}arlo on graphical processing units. \emph{Journal of Chemical Theory and Computation} \textbf{2018}, \emph{14}, 4109--4121\relax
\mciteBstWouldAddEndPuncttrue
\mciteSetBstMidEndSepPunct{\mcitedefaultmidpunct}
{\mcitedefaultendpunct}{\mcitedefaultseppunct}\relax
\EndOfBibitem
\bibitem[Malone \latin{et~al.}(2020)Malone, Zhang, and Morales]{malone2020accelerating}
Malone,~F.~D.; Zhang,~S.; Morales,~M.~A. Accelerating auxiliary-field quantum Monte Carlo simulations of solids with graphical processing units. \emph{Journal of Chemical Theory and Computation} \textbf{2020}, \emph{16}, 4286--4297\relax
\mciteBstWouldAddEndPuncttrue
\mciteSetBstMidEndSepPunct{\mcitedefaultmidpunct}
{\mcitedefaultendpunct}{\mcitedefaultseppunct}\relax
\EndOfBibitem
\bibitem[Jiang \latin{et~al.}(2024)Jiang, Baumgarten, Loos, Mahajan, Scemama, Ung, Zhang, Malone, and Lee]{jiang2024improved}
Jiang,~T.; Baumgarten,~M.~K.; Loos,~P.-F.; Mahajan,~A.; Scemama,~A.; Ung,~S.~F.; Zhang,~J.; Malone,~F.~D.; Lee,~J. Improved modularity and new features in ipie: Toward even larger AFQMC calculations on CPUs and GPUs at zero and finite temperatures. \emph{The Journal of Chemical Physics} \textbf{2024}, \emph{161}\relax
\mciteBstWouldAddEndPuncttrue
\mciteSetBstMidEndSepPunct{\mcitedefaultmidpunct}
{\mcitedefaultendpunct}{\mcitedefaultseppunct}\relax
\EndOfBibitem
\bibitem[Motta and Zhang(2017)Motta, and Zhang]{motta2017computation}
Motta,~M.; Zhang,~S. Computation of ground-state properties in molecular systems: Back-propagation with auxiliary-field quantum Monte Carlo. \emph{Journal of Chemical Theory and Computation} \textbf{2017}, \emph{13}, 5367--5378\relax
\mciteBstWouldAddEndPuncttrue
\mciteSetBstMidEndSepPunct{\mcitedefaultmidpunct}
{\mcitedefaultendpunct}{\mcitedefaultseppunct}\relax
\EndOfBibitem
\bibitem[Motta and Zhang(2018)Motta, and Zhang]{motta2018communication}
Motta,~M.; Zhang,~S. Communication: Calculation of interatomic forces and optimization of molecular geometry with auxiliary-field quantum Monte Carlo. \emph{The Journal of Chemical Physics} \textbf{2018}, \emph{148}\relax
\mciteBstWouldAddEndPuncttrue
\mciteSetBstMidEndSepPunct{\mcitedefaultmidpunct}
{\mcitedefaultendpunct}{\mcitedefaultseppunct}\relax
\EndOfBibitem
\bibitem[Mahajan \latin{et~al.}(2023)Mahajan, Kurian, Lee, Reichman, and Sharma]{mahajan2023response}
Mahajan,~A.; Kurian,~J.~S.; Lee,~J.; Reichman,~D.~R.; Sharma,~S. Response properties in phaseless auxiliary field quantum Monte Carlo. \emph{The Journal of Chemical Physics} \textbf{2023}, \emph{159}\relax
\mciteBstWouldAddEndPuncttrue
\mciteSetBstMidEndSepPunct{\mcitedefaultmidpunct}
{\mcitedefaultendpunct}{\mcitedefaultseppunct}\relax
\EndOfBibitem
\bibitem[Ma \latin{et~al.}(2013)Ma, Zhang, and Krakauer]{ma_excited_2013}
Ma,~F.; Zhang,~S.; Krakauer,~H. Excited state calculations in solids by auxiliary-field quantum {Monte} {Carlo}. \emph{New Journal of Physics} \textbf{2013}, \emph{15}, 093017\relax
\mciteBstWouldAddEndPuncttrue
\mciteSetBstMidEndSepPunct{\mcitedefaultmidpunct}
{\mcitedefaultendpunct}{\mcitedefaultseppunct}\relax
\EndOfBibitem
\bibitem[Shee \latin{et~al.}(2019)Shee, Rudshteyn, Arthur, Zhang, Reichman, and Friesner]{shee2019achieving}
Shee,~J.; Rudshteyn,~B.; Arthur,~E.~J.; Zhang,~S.; Reichman,~D.~R.; Friesner,~R.~A. On achieving high accuracy in quantum chemical calculations of $3d$ transition metal-containing systems: a comparison of auxiliary-field quantum Monte Carlo with coupled cluster, density functional theory, and experiment for diatomic molecules. \emph{Journal of Chemical Theory and Computation} \textbf{2019}, \emph{15}, 2346--2358\relax
\mciteBstWouldAddEndPuncttrue
\mciteSetBstMidEndSepPunct{\mcitedefaultmidpunct}
{\mcitedefaultendpunct}{\mcitedefaultseppunct}\relax
\EndOfBibitem
\bibitem[Mahajan \latin{et~al.}(2024)Mahajan, Thorpe, Kurian, Reichman, Matthews, and Sharma]{mahajan2024beyond}
Mahajan,~A.; Thorpe,~J.~H.; Kurian,~J.~S.; Reichman,~D.~R.; Matthews,~D.~A.; Sharma,~S. Beyond CCSD (T) accuracy at lower scaling with auxiliary field quantum Monte Carlo. \emph{Journal of Chemical Theory and Computation} \textbf{2024}, \relax
\mciteBstWouldAddEndPunctfalse
\mciteSetBstMidEndSepPunct{\mcitedefaultmidpunct}
{}{\mcitedefaultseppunct}\relax
\EndOfBibitem
\bibitem[Rudshteyn \latin{et~al.}(2020)Rudshteyn, Coskun, Weber, Arthur, Zhang, Reichman, Friesner, and Shee]{rudshteyn2020predicting}
Rudshteyn,~B.; Coskun,~D.; Weber,~J.~L.; Arthur,~E.~J.; Zhang,~S.; Reichman,~D.~R.; Friesner,~R.~A.; Shee,~J. Predicting ligand-dissociation energies of $3d$ coordination complexes with auxiliary-field quantum {M}onte {C}arlo. \emph{Journal of Chemical Theory and Computation} \textbf{2020}, \emph{16}, 3041--3054\relax
\mciteBstWouldAddEndPuncttrue
\mciteSetBstMidEndSepPunct{\mcitedefaultmidpunct}
{\mcitedefaultendpunct}{\mcitedefaultseppunct}\relax
\EndOfBibitem
\bibitem[Rudshteyn \latin{et~al.}(2022)Rudshteyn, Weber, Coskun, Devlaminck, Zhang, Reichman, Shee, and Friesner]{rudshteyn2022calculation}
Rudshteyn,~B.; Weber,~J.~L.; Coskun,~D.; Devlaminck,~P.~A.; Zhang,~S.; Reichman,~D.~R.; Shee,~J.; Friesner,~R.~A. Calculation of metallocene ionization potentials via auxiliary field quantum Monte Carlo: Toward benchmark quantum chemistry for transition metals. \emph{Journal of Chemical Theory and Computation} \textbf{2022}, \emph{18}, 2845--2862\relax
\mciteBstWouldAddEndPuncttrue
\mciteSetBstMidEndSepPunct{\mcitedefaultmidpunct}
{\mcitedefaultendpunct}{\mcitedefaultseppunct}\relax
\EndOfBibitem
\bibitem[Shee \latin{et~al.}(2019)Shee, Arthur, Zhang, Reichman, and Friesner]{shee2019singlet}
Shee,~J.; Arthur,~E.~J.; Zhang,~S.; Reichman,~D.~R.; Friesner,~R.~A. Singlet--triplet energy gaps of organic biradicals and polyacenes with auxiliary-field quantum {M}onte {C}arlo. \emph{Journal of Chemical Theory and Computation} \textbf{2019}, \emph{15}, 4924--4932\relax
\mciteBstWouldAddEndPuncttrue
\mciteSetBstMidEndSepPunct{\mcitedefaultmidpunct}
{\mcitedefaultendpunct}{\mcitedefaultseppunct}\relax
\EndOfBibitem
\bibitem[Weber \latin{et~al.}(2021)Weber, Churchill, Jockusch, Arthur, Pun, Zhang, Friesner, Campos, Reichman, and Shee]{weber2021silico}
Weber,~J.~L.; Churchill,~E.~M.; Jockusch,~S.; Arthur,~E.~J.; Pun,~A.~B.; Zhang,~S.; Friesner,~R.~A.; Campos,~L.~M.; Reichman,~D.~R.; Shee,~J. In silico prediction of annihilators for triplet--triplet annihilation upconversion via auxiliary-field quantum {M}onte {C}arlo. \emph{Chemical Science} \textbf{2021}, \emph{12}, 1068--1079\relax
\mciteBstWouldAddEndPuncttrue
\mciteSetBstMidEndSepPunct{\mcitedefaultmidpunct}
{\mcitedefaultendpunct}{\mcitedefaultseppunct}\relax
\EndOfBibitem
\bibitem[Lee \latin{et~al.}(2020)Lee, Malone, and Morales]{lee2020utilizing}
Lee,~J.; Malone,~F.~D.; Morales,~M.~A. Utilizing essential symmetry breaking in auxiliary-field quantum Monte Carlo: Application to the spin gaps of the C$_{36}$ fullerene and an iron porphyrin model complex. \emph{Journal of Chemical Theory and Computation} \textbf{2020}, \emph{16}, 3019--3027\relax
\mciteBstWouldAddEndPuncttrue
\mciteSetBstMidEndSepPunct{\mcitedefaultmidpunct}
{\mcitedefaultendpunct}{\mcitedefaultseppunct}\relax
\EndOfBibitem
\bibitem[Hao \latin{et~al.}(2018)Hao, Shee, Upadhyay, Ataca, Jordan, and Rubenstein]{hao2018accurate}
Hao,~H.; Shee,~J.; Upadhyay,~S.; Ataca,~C.; Jordan,~K.~D.; Rubenstein,~B.~M. Accurate predictions of electron binding energies of dipole-bound anions via quantum {M}onte {C}arlo methods. \emph{The Journal of Physical Chemistry Letters} \textbf{2018}, \emph{9}, 6185--6190\relax
\mciteBstWouldAddEndPuncttrue
\mciteSetBstMidEndSepPunct{\mcitedefaultmidpunct}
{\mcitedefaultendpunct}{\mcitedefaultseppunct}\relax
\EndOfBibitem
\bibitem[Upadhyay \latin{et~al.}(2020)Upadhyay, Dumi, Shee, and Jordan]{upadhyay2020role}
Upadhyay,~S.; Dumi,~A.; Shee,~J.; Jordan,~K.~D. The role of high-order electron correlation effects in a model system for non-valence correlation-bound anions. \emph{The Journal of Chemical Physics} \textbf{2020}, \emph{153}, 224118\relax
\mciteBstWouldAddEndPuncttrue
\mciteSetBstMidEndSepPunct{\mcitedefaultmidpunct}
{\mcitedefaultendpunct}{\mcitedefaultseppunct}\relax
\EndOfBibitem
\bibitem[Sukurma \latin{et~al.}(2023)Sukurma, Schlipf, Humer, Taheridehkordi, and Kresse]{sukurma2023benchmark}
Sukurma,~Z.; Schlipf,~M.; Humer,~M.; Taheridehkordi,~A.; Kresse,~G. Benchmark phaseless auxiliary-field quantum Monte Carlo method for small molecules. \emph{Journal of Chemical Theory and Computation} \textbf{2023}, \emph{19}, 4921--4934\relax
\mciteBstWouldAddEndPuncttrue
\mciteSetBstMidEndSepPunct{\mcitedefaultmidpunct}
{\mcitedefaultendpunct}{\mcitedefaultseppunct}\relax
\EndOfBibitem
\bibitem[Ganoe and Shee(2024)Ganoe, and Shee]{ganoe_notion_2024}
Ganoe,~B.; Shee,~J. On the notion of strong correlation in electronic structure theory. \emph{Faraday Discussions} \textbf{2024}, 10.1039.D4FD00066H\relax
\mciteBstWouldAddEndPuncttrue
\mciteSetBstMidEndSepPunct{\mcitedefaultmidpunct}
{\mcitedefaultendpunct}{\mcitedefaultseppunct}\relax
\EndOfBibitem
\bibitem[Holmes \latin{et~al.}(2016)Holmes, Tubman, and Umrigar]{holmes2016heat}
Holmes,~A.~A.; Tubman,~N.~M.; Umrigar,~C. Heat-bath configuration interaction: An efficient selected configuration interaction algorithm inspired by heat-bath sampling. \emph{Journal of Chemical Theory and Computation} \textbf{2016}, \emph{12}, 3674--3680\relax
\mciteBstWouldAddEndPuncttrue
\mciteSetBstMidEndSepPunct{\mcitedefaultmidpunct}
{\mcitedefaultendpunct}{\mcitedefaultseppunct}\relax
\EndOfBibitem
\bibitem[Neugebauer \latin{et~al.}(2023)Neugebauer, Vuong, Weber, Friesner, Shee, and Hansen]{neugebauer2023toward}
Neugebauer,~H.; Vuong,~H.~T.; Weber,~J.~L.; Friesner,~R.~A.; Shee,~J.; Hansen,~A. Toward benchmark-quality ab initio predictions for $3d$ transition metal electrocatalysts: a comparison of {CCSD(T)} and {ph-AFQMC}. \emph{Journal of Chemical Theory and Computation} \textbf{2023}, \emph{19}, 6208--6225\relax
\mciteBstWouldAddEndPuncttrue
\mciteSetBstMidEndSepPunct{\mcitedefaultmidpunct}
{\mcitedefaultendpunct}{\mcitedefaultseppunct}\relax
\EndOfBibitem
\bibitem[Mahajan \latin{et~al.}(2022)Mahajan, Lee, and Sharma]{mahajan2022selected}
Mahajan,~A.; Lee,~J.; Sharma,~S. Selected configuration interaction wave functions in phaseless auxiliary field quantum Monte Carlo. \emph{The Journal of Chemical Physics} \textbf{2022}, \emph{156}\relax
\mciteBstWouldAddEndPuncttrue
\mciteSetBstMidEndSepPunct{\mcitedefaultmidpunct}
{\mcitedefaultendpunct}{\mcitedefaultseppunct}\relax
\EndOfBibitem
\bibitem[Jiang \latin{et~al.}(2025)Jiang, O'Gorman, Mahajan, and Lee]{jiang2025unbiasing}
Jiang,~T.; O'Gorman,~B.; Mahajan,~A.; Lee,~J. Unbiasing fermionic auxiliary-field quantum Monte Carlo with matrix product state trial wavefunctions. \emph{Physical Review Research} \textbf{2025}, \emph{7}, 013038\relax
\mciteBstWouldAddEndPuncttrue
\mciteSetBstMidEndSepPunct{\mcitedefaultmidpunct}
{\mcitedefaultendpunct}{\mcitedefaultseppunct}\relax
\EndOfBibitem
\bibitem[Danilov \latin{et~al.}(2024)Danilov, Ganoe, Munyi, and Shee]{danilov_capturing_2024}
Danilov,~D.; Ganoe,~B.; Munyi,~M.; Shee,~J. Capturing strong correlation in molecules with phaseless auxiliary-field quantum {Monte} {Carlo} using generalized {Hartree} {Fock} trial wavefunctions. 2024; \url{https://chemrxiv.org/engage/chemrxiv/article-details/66ed97f1cec5d6c142ab8159}\relax
\mciteBstWouldAddEndPuncttrue
\mciteSetBstMidEndSepPunct{\mcitedefaultmidpunct}
{\mcitedefaultendpunct}{\mcitedefaultseppunct}\relax
\EndOfBibitem
\bibitem[{Qiskit contributors}(2023)]{Qiskit}
{Qiskit contributors} Qiskit: An Open-source Framework for Quantum Computing. 2023; \url{10.5281/zenodo.2573505}\relax
\mciteBstWouldAddEndPuncttrue
\mciteSetBstMidEndSepPunct{\mcitedefaultmidpunct}
{\mcitedefaultendpunct}{\mcitedefaultseppunct}\relax
\EndOfBibitem
\bibitem[Johnson \latin{et~al.}(2024)Johnson, Barison, Fuller, Garrison, Glick, Saki, Mezzacapo, Robledo-Moreno, Rossmannek, Schweigert, Sitdikov, and Sung]{sqd_addon}
Johnson,~C.; Barison,~S.; Fuller,~B.; Garrison,~J.~R.; Glick,~J.~R.; Saki,~A.~A.; Mezzacapo,~A.; Robledo-Moreno,~J.; Rossmannek,~M.; Schweigert,~P.; Sitdikov,~I.; Sung,~K.~J. {Qiskit addon: sample-based quantum diagonalization}. \url{https://github.com/Qiskit/qiskit-addon-sqd}, 2024\relax
\mciteBstWouldAddEndPuncttrue
\mciteSetBstMidEndSepPunct{\mcitedefaultmidpunct}
{\mcitedefaultendpunct}{\mcitedefaultseppunct}\relax
\EndOfBibitem
\bibitem[{The ffsim developers}(2024)]{ffsim}
{The ffsim developers} {ffsim: Faster simulations of fermionic quantum circuits}. 2024; \url{https://github.com/qiskit-community/ffsim}\relax
\mciteBstWouldAddEndPuncttrue
\mciteSetBstMidEndSepPunct{\mcitedefaultmidpunct}
{\mcitedefaultendpunct}{\mcitedefaultseppunct}\relax
\EndOfBibitem
\bibitem[Virtanen \latin{et~al.}(2020)Virtanen, Gommers, Oliphant, Haberland, Reddy, Cournapeau, Burovski, Peterson, Weckesser, Bright, {van der Walt}, Brett, Wilson, Millman, Mayorov, Nelson, Jones, Kern, Larson, Carey, Polat, Feng, Moore, {VanderPlas}, Laxalde, Perktold, Cimrman, Henriksen, Quintero, Harris, Archibald, Ribeiro, Pedregosa, {van Mulbregt}, and {SciPy 1.0 Contributors}]{scipy}
Virtanen,~P.; Gommers,~R.; Oliphant,~T.~E.; Haberland,~M.; Reddy,~T.; Cournapeau,~D.; Burovski,~E.; Peterson,~P.; Weckesser,~W.; Bright,~J.; {van der Walt},~S.~J.; Brett,~M.; Wilson,~J.; Millman,~K.~J.; Mayorov,~N.; Nelson,~A. R.~J.; Jones,~E.; Kern,~R.; Larson,~E.; Carey,~C.~J.; Polat,~{\.I}.; Feng,~Y.; Moore,~E.~W.; {VanderPlas},~J.; Laxalde,~D.; Perktold,~J.; Cimrman,~R.; Henriksen,~I.; Quintero,~E.~A.; Harris,~C.~R.; Archibald,~A.~M.; Ribeiro,~A.~H.; Pedregosa,~F.; {van Mulbregt},~P.; {SciPy 1.0 Contributors} {{SciPy} 1.0: Fundamental Algorithms for Scientific Computing in Python}. \emph{Nature Methods} \textbf{2020}, \emph{17}, 261--272\relax
\mciteBstWouldAddEndPuncttrue
\mciteSetBstMidEndSepPunct{\mcitedefaultmidpunct}
{\mcitedefaultendpunct}{\mcitedefaultseppunct}\relax
\EndOfBibitem
\bibitem[Flyvbjerg and Petersen(1989)Flyvbjerg, and Petersen]{reblockana}
Flyvbjerg,~H.; Petersen,~H.~G. Error estimates on averages of correlated data. \emph{The Journal of Chemical Physics} \textbf{1989}, \emph{91}, 461--466\relax
\mciteBstWouldAddEndPuncttrue
\mciteSetBstMidEndSepPunct{\mcitedefaultmidpunct}
{\mcitedefaultendpunct}{\mcitedefaultseppunct}\relax
\EndOfBibitem
\bibitem[Sun \latin{et~al.}(2018)Sun, Berkelbach, Blunt, Booth, Guo, Li, Liu, McClain, Sayfutyarova, Sharma, \latin{et~al.} others]{sun2018pyscf}
Sun,~Q.; Berkelbach,~T.~C.; Blunt,~N.~S.; Booth,~G.~H.; Guo,~S.; Li,~Z.; Liu,~J.; McClain,~J.~D.; Sayfutyarova,~E.~R.; Sharma,~S.; others {P}y{SCF}: the Python-based simulations of chemistry framework. \emph{Wiley Interdisciplinary Reviews: Computational Molecular Science} \textbf{2018}, \emph{8}, e1340\relax
\mciteBstWouldAddEndPuncttrue
\mciteSetBstMidEndSepPunct{\mcitedefaultmidpunct}
{\mcitedefaultendpunct}{\mcitedefaultseppunct}\relax
\EndOfBibitem
\bibitem[Sun \latin{et~al.}(2020)Sun, Zhang, Banerjee, Bao, Barbry, Blunt, Bogdanov, Booth, Chen, Cui, \latin{et~al.} others]{sun2020recent}
Sun,~Q.; Zhang,~X.; Banerjee,~S.; Bao,~P.; Barbry,~M.; Blunt,~N.~S.; Bogdanov,~N.~A.; Booth,~G.~H.; Chen,~J.; Cui,~Z.-H.; others Recent developments in the {PySCF} program package. \emph{The Journal of Chemical Physics} \textbf{2020}, \emph{153}, 024109\relax
\mciteBstWouldAddEndPuncttrue
\mciteSetBstMidEndSepPunct{\mcitedefaultmidpunct}
{\mcitedefaultendpunct}{\mcitedefaultseppunct}\relax
\EndOfBibitem
\bibitem[Sharma \latin{et~al.}(2014)Sharma, Sivalingam, Neese, and Chan]{sharma2014low}
Sharma,~S.; Sivalingam,~K.; Neese,~F.; Chan,~G. K.-L. Low-energy spectrum of iron-sulfur clusters directly from many-particle quantum mechanics. \emph{Nature Chemistry} \textbf{2014}, \emph{6}, 927--933\relax
\mciteBstWouldAddEndPuncttrue
\mciteSetBstMidEndSepPunct{\mcitedefaultmidpunct}
{\mcitedefaultendpunct}{\mcitedefaultseppunct}\relax
\EndOfBibitem
\bibitem[Kashima and Imada(2001)Kashima, and Imada]{kashima2001path}
Kashima,~T.; Imada,~M. Path-integral renormalization group method for numerical study on ground states of strongly correlated electronic systems. \emph{Journal of the Physical Society of Japan} \textbf{2001}, \emph{70}, 2287--2299\relax
\mciteBstWouldAddEndPuncttrue
\mciteSetBstMidEndSepPunct{\mcitedefaultmidpunct}
{\mcitedefaultendpunct}{\mcitedefaultseppunct}\relax
\EndOfBibitem
\bibitem[Holmes \latin{et~al.}(2016)Holmes, Tubman, and Umrigar]{dice1}
Holmes,~A.~A.; Tubman,~N.~M.; Umrigar,~C.~J. Heat-bath configuration interaction: An efficient selected configuration interaction algorithm inspired by heat-bath sampling. \emph{J. Chem. Theory Comput.} \textbf{2016}, \emph{12}, 3674--3680\relax
\mciteBstWouldAddEndPuncttrue
\mciteSetBstMidEndSepPunct{\mcitedefaultmidpunct}
{\mcitedefaultendpunct}{\mcitedefaultseppunct}\relax
\EndOfBibitem
\bibitem[Sharma \latin{et~al.}(2017)Sharma, Holmes, Jeanmairet, Alavi, and Umrigar]{dice2}
Sharma,~S.; Holmes,~A.~A.; Jeanmairet,~G.; Alavi,~A.; Umrigar,~C.~J. Semistochastic heat-bath configuration interaction method: Selected configuration interaction with semistochastic perturbation theory. \emph{J. Chem. Theory Comput.} \textbf{2017}, \emph{13}, 1595--1604\relax
\mciteBstWouldAddEndPuncttrue
\mciteSetBstMidEndSepPunct{\mcitedefaultmidpunct}
{\mcitedefaultendpunct}{\mcitedefaultseppunct}\relax
\EndOfBibitem
\bibitem[Li and Chan(2017)Li, and Chan]{li_spin-projected_2017}
Li,~Z.; Chan,~G. K.-L. Spin-{Projected} {Matrix} {Product} {States}: {Versatile} {Tool} for {Strongly} {Correlated} {Systems}. \emph{Journal of Chemical Theory and Computation} \textbf{2017}, \emph{13}, 2681--2695\relax
\mciteBstWouldAddEndPuncttrue
\mciteSetBstMidEndSepPunct{\mcitedefaultmidpunct}
{\mcitedefaultendpunct}{\mcitedefaultseppunct}\relax
\EndOfBibitem
\bibitem[Huggins \latin{et~al.}(2021)Huggins, McClean, Rubin, Jiang, Wiebe, Whaley, and Babbush]{huggins2021efficient}
Huggins,~W.~J.; McClean,~J.~R.; Rubin,~N.~C.; Jiang,~Z.; Wiebe,~N.; Whaley,~K.~B.; Babbush,~R. Efficient and noise resilient measurements for quantum chemistry on near-term quantum computers. \emph{npj Quantum Information} \textbf{2021}, \emph{7}, 23\relax
\mciteBstWouldAddEndPuncttrue
\mciteSetBstMidEndSepPunct{\mcitedefaultmidpunct}
{\mcitedefaultendpunct}{\mcitedefaultseppunct}\relax
\EndOfBibitem
\bibitem[Wang \latin{et~al.}(2021)Wang, Fontana, Cerezo, Sharma, Sone, Cincio, and Coles]{wang2021noise}
Wang,~S.; Fontana,~E.; Cerezo,~M.; Sharma,~K.; Sone,~A.; Cincio,~L.; Coles,~P.~J. Noise-induced barren plateaus in variational quantum algorithms. \emph{Nature communications} \textbf{2021}, \emph{12}, 6961\relax
\mciteBstWouldAddEndPuncttrue
\mciteSetBstMidEndSepPunct{\mcitedefaultmidpunct}
{\mcitedefaultendpunct}{\mcitedefaultseppunct}\relax
\EndOfBibitem
\bibitem[Patel \latin{et~al.}(2025)Patel, Jayakumar, Yen, and Izmaylov]{patel2025quantum}
Patel,~S.; Jayakumar,~P.; Yen,~T.-C.; Izmaylov,~A.~F. Quantum Measurement for Quantum Chemistry on a Quantum Computer. \emph{arXiv preprint arXiv:2501.14968} \textbf{2025}, \relax
\mciteBstWouldAddEndPunctfalse
\mciteSetBstMidEndSepPunct{\mcitedefaultmidpunct}
{}{\mcitedefaultseppunct}\relax
\EndOfBibitem
\bibitem[Huang \latin{et~al.}(2024)Huang, Guo, Pham, and Lv]{huang2024gpu}
Huang,~Y.; Guo,~Z.; Pham,~H.~Q.; Lv,~D. GPU-accelerated Auxiliary-field quantum Monte Carlo with multi-Slater determinant trial states. \emph{arXiv preprint arXiv:2406.08314} \textbf{2024}, \relax
\mciteBstWouldAddEndPunctfalse
\mciteSetBstMidEndSepPunct{\mcitedefaultmidpunct}
{}{\mcitedefaultseppunct}\relax
\EndOfBibitem
\bibitem[Mikkelsen and Nakagawa(2025)Mikkelsen, and Nakagawa]{mikkelsen2025quantumselectedconfigurationinteractiontimeevolved}
Mikkelsen,~M.; Nakagawa,~Y.~O. Quantum-selected configuration interaction with time-evolved state. 2025; \url{https://arxiv.org/abs/2412.13839}\relax
\mciteBstWouldAddEndPuncttrue
\mciteSetBstMidEndSepPunct{\mcitedefaultmidpunct}
{\mcitedefaultendpunct}{\mcitedefaultseppunct}\relax
\EndOfBibitem
\bibitem[Sugisaki \latin{et~al.}(2025)Sugisaki, Kanno, Itoko, Sakuma, and Yamamoto]{sugisaki2025hamiltoniansimulationbasedquantumselectedconfiguration}
Sugisaki,~K.; Kanno,~S.; Itoko,~T.; Sakuma,~R.; Yamamoto,~N. Hamiltonian simulation-based quantum-selected configuration interaction for large-scale electronic structure calculations with a quantum computer. 2025; \url{https://arxiv.org/abs/2412.07218}\relax
\mciteBstWouldAddEndPuncttrue
\mciteSetBstMidEndSepPunct{\mcitedefaultmidpunct}
{\mcitedefaultendpunct}{\mcitedefaultseppunct}\relax
\EndOfBibitem
\bibitem[Yu \latin{et~al.}(2025)Yu, Moreno, Iosue, Bertels, Claudino, Fuller, Groszkowski, Humble, Jurcevic, Kirby, \latin{et~al.} others]{yu2025quantum}
Yu,~J.; Moreno,~J.~R.; Iosue,~J.~T.; Bertels,~L.; Claudino,~D.; Fuller,~B.; Groszkowski,~P.; Humble,~T.~S.; Jurcevic,~P.; Kirby,~W.; others Quantum-centric algorithm for sample-based Krylov diagonalization. \emph{arXiv preprint arXiv:2501.09702} \textbf{2025}, \relax
\mciteBstWouldAddEndPunctfalse
\mciteSetBstMidEndSepPunct{\mcitedefaultmidpunct}
{}{\mcitedefaultseppunct}\relax
\EndOfBibitem
\bibitem[Yoshida \latin{et~al.}(2025)Yoshida, Erhart, Murokoshi, Nakagawa, Mori, Miyanaga, Mori, and Mizukami]{yoshida2025auxiliary}
Yoshida,~Y.; Erhart,~L.; Murokoshi,~T.; Nakagawa,~R.; Mori,~C.; Miyanaga,~T.; Mori,~T.; Mizukami,~W. Auxiliary-field quantum Monte Carlo method with quantum selected configuration interaction. \emph{arXiv preprint arXiv:2502.21081} \textbf{2025}, \relax
\mciteBstWouldAddEndPunctfalse
\mciteSetBstMidEndSepPunct{\mcitedefaultmidpunct}
{}{\mcitedefaultseppunct}\relax
\EndOfBibitem
\end{mcitethebibliography}
\end{document}